\documentclass[11pt,fleqn]{article}
	
	\usepackage{amssymb,amsmath}
	\usepackage{natbib}
	\usepackage{vmargin,setspace,verbatim}
	\usepackage{graphicx,graphics,color,tikz,tikz-network}
	\usepackage{microtype}
	
	\usetikzlibrary{arrows.meta, calc, fit, backgrounds, decorations.pathreplacing}
	
\usepackage{libertine}
\usepackage[libertine]{newtxmath}

\definecolor{darkgreen}{rgb}{0,0.2,0}
\definecolor{darkred}{rgb}{0.3,0,0}
\definecolor{lyellow}{rgb}{1,1,0.7}
\definecolor{lblue}{rgb}{0.7,0.9,1}
	
\usepackage[colorlinks=true, pdfstartview=FitV, linkcolor=darkgreen, citecolor=darkred, urlcolor=black]{hyperref}

\newcounter{llst}
\newenvironment{abet}{\begin{list}{\rm (\alph{llst})}{\usecounter{llst}
		\setlength{\itemindent}{0em} \setlength{\leftmargin}{3em}
		\setlength{\labelwidth}{2em} \setlength{\labelsep}{1em}}}{\end{list}}

\newcounter{llist}
\newenvironment{numm}{\begin{list}{\rm (\roman{llist})}{\usecounter{llist}
	\setlength{\itemindent}{0em} \setlength{\leftmargin}{3.5em}
	\setlength{\labelwidth}{2.5em} \setlength{\labelsep}{1em}}}{\end{list}}

\newtheorem{theorem}{Theorem}[section]

\newtheorem{corollary}[theorem]{Corollary}

\newtheorem{definition}[theorem]{Definition}
\newtheorem{expl}[theorem]{Example}

\newtheorem{lemma}[theorem]{Lemma}

\newtheorem{proposition}[theorem]{Proposition}
\newtheorem{remark}[theorem]{Remark}

\newtheorem{dscrpt}[theorem]{Description}

\newcounter{axiomatiser}

\newenvironment{example}{\begin{expl} \rm}{\hfill $\blacklozenge$
\end{expl}}

\setpapersize{A4}
\setmarginsrb{80pt}{40pt}{80pt}{60pt}{12pt}{10pt}{12pt}{30pt}

\newcommand{\rx}[1]{%
  \draw[red, line width=1.3pt] ($#1+(-0.16,-0.16)$) -- ($#1+(0.16,0.16)$);
  \draw[red, line width=1.3pt] ($#1+(-0.16, 0.16)$) -- ($#1+(0.16,-0.16)$);
}

\begin{document}

\title{\textbf{Structured Production Systems: Viability}\thanks{We thank Achille Basile for his support and thoughtful comments during the development of this paper. We are also grateful to Chiara Donnini and the audience at the workshop \lq \lq Strategic Interactions and General Equilibrium XIV:  Theories and Applications\rq \rq  at University Paris Nanterre. A special thanks goes to the two anonymous referees for their careful reading and valuable suggestions. }}

\author{Robert P.~Gilles\thanks{Department of Economics, The Queen's University of Belfast, Riddel Hall, 185 Stranmillis Road, Belfast, BT9~5EE, UK. \textsf{Email: r.gilles@qub.ac.uk}} \and Marialaura Pesce\thanks{\textbf{Corresponding author.}  Dipartimento di Scienze Economiche e Statistiche and CSEF, Universit\`a di Napoli Federico II, Complesso Universitario Monte Sant'Angelo Via Cintia, 80126 Napoli, Italy. \textsf{Email: marialaura.pesce@unina.it} M.~Pesce acknowledges the European Union -- Next Generation EU, in the framework of the GRINS -- Growing Resilient, INclusive and Sustainable project (GRINS PE00000018 - CUP E63C22002140007) and financial support of the Italian Ministry of University and Research (MUR) under the PRIN 2022 (HLPMKN). }}

\date{January 2026 \\ Revised: June 2026}


\maketitle

\begin{singlespace}
	
\begin{abstract}
\noindent
This paper introduces a novel framework for analysing equilibrium in structured production systems that incorporate a static social division of labour, distinguishing between consumption goods traded in competitive markets and intermediate goods exchanged through bilateral relationships. We develop the concept of viability---the requirement that all producers earn positive incomes---as a foundational equilibrium prerequisite.

Our main theorem establishes that a structured production system is viable if and only if it is coherent---admitting no circular conversion processes that yield no net output---which in turn is equivalent to the non-singularity of its matrix representation. We further investigate completely viable systems, in which viable prices exist for all consumption good price vectors. We show that complete viability stands exactly between two input restrictions: it is guaranteed in coherent systems where no consumption good is used as an input in production, and it requires that no consumption good is used as an input in the production of another consumption good. The analysis thus reveals fundamental relationships between the architectural design of production systems and their economic sustainability.

Our framework also contributes to the literature on the existence of positive output price systems and the Hawkins-Simon condition in input-output analysis.
\end{abstract}

\medskip
\begin{description}
	\item[JEL classification:] C62, C67, D51
	\item[Keywords:] Structured production system; Social division of labour; Viability; Viable price system
\end{description}

\end{singlespace}

\thispagestyle{empty}

\newpage

\setcounter{page}{1} \pagenumbering{arabic}

\section{Supply chains and structured production}

Modern economies are characterised by complex production networks where goods flow through multiple stages of transformation before reaching final consumers. This paper introduces a novel framework for analysing equilibrium in such structured production systems, departing fundamentally from traditional approaches by distinguishing between two categories of economic goods: consumption goods traded in competitive markets and intermediate goods exchanged through bilateral relationships within production chains. This distinction, while intuitive from an empirical standpoint, has profound implications for equilibrium theory and requires a reconceptualisation of how prices emerge and sustain productive activity.

\paragraph{Differentiation of economic goods}

Our framework recognises that economic goods serve fundamentally different roles in production systems. Consumption goods represent final outputs brought to competitive markets where prices emerge through supply and demand. These markets operate according to standard competitive principles, with prices serving as signals that coordinate the decisions of numerous anonymous agents. In contrast, intermediate goods function as inputs within production chains, traded bilaterally between specific producers according to technical requirements rather than through anonymous market mechanisms.

This distinction challenges the unified treatment of goods in classical general equilibrium theory---set out by \citet{ArrowDebreu1954} and \citet{Debreu1959}---and input-output analysis \citep{Leontief1936,Leontief1941}. While Arrow-Debreu models typically abstract from production chains through a model based on production sets, and Leontief models emphasise technical coefficients without explicit price formation mechanisms, our approach captures the dual nature of modern production systems where market and non-market, network exchanges coexist. The separation reflects empirical realities where intermediate goods often involve relationship-specific investments, technical specifications, and quality requirements that preclude anonymous market exchange. Their prices emerge through bargaining between connected producers rather than competitive forces. This institutional reality necessitates a theoretical framework that accommodates both competitive pricing for final goods and negotiated pricing for intermediates.

\paragraph{A three-stage equilibrium conception}

The comprehensive analysis of equilibrium in structured production systems requires a carefully sequenced theoretical conception. This paper initiates a three-stage equilibrium concept, with each stage addressing increasingly complex aspects of price determination and equilibrium.

The first stage, developed in this paper, establishes the concept of viability---the requirement that all producers earn positive incomes at prevailing prices. Without viability, producers would exit the system, disrupting production chains and preventing the realisation of any equilibrium. We characterise the structural properties of production systems that guarantee the existence of viable price systems, establishing connections between the architecture or design of the production system and economic sustainability. This analysis relies on the use of matrix theory, in particular the Hawkins-Simon condition \citep{HawkinsSimon1949} and McKenzie's condition of a quasi-dominant diagonal in a $\mathcal Z$-matrix \citep{McKenzie1960}. This foundational analysis identifies when price systems can sustain all productive activities, a prerequisite for any meaningful equilibrium concept.

The second stage, to be developed in a companion paper, introduces a general equilibrium concept that combines viability with market clearing in consumption good markets. This equilibrium notion must reconcile two distinct coordination mechanisms: competitive markets for consumption goods and bilateral bargaining relationships for intermediate goods. The challenge lies in proving existence while respecting the institutional constraint that intermediate goods lack organised markets.

The third stage will endogenise intermediate good prices through explicit bargaining mechanisms. We envision two complementary approaches. First, reference groups of producers with similar economic standing may negotiate on equal footing, leading to income equalisation within groups. Alternatively, a Nash bargaining framework can model situations where producers possess differential bargaining power, yielding specific income distributions determined by network position and economic leverage.

\subsection{Relationship to existing literature}

\paragraph{Production networks and structural analysis}

Recent advances in network economics have renewed interest in production structures. \citet{Carvalho2019} provide a primer on this literature, organised around a benchmark static general equilibrium model of production networks in which each good may serve simultaneously as a final consumption good and as an intermediate input, with all goods priced in competitive markets. In that framework, equilibrium prices and the transmission of shocks are governed by the Leontief inverse $(I-A)^{-1}$, whose existence and non-negativity follow from the input-output matrix $A$ having a spectral radius below one---a property guaranteed at the outset by constant returns to scale together with a positive labour share, and equivalent to $I-A$ being a non-singular $\mathcal Z$-matrix satisfying the Hawkins-Simon condition that also underpins our analysis. 

Our framework departs from this unified treatment in two related ways: we distinguish competitively priced consumption goods from bilaterally bargained intermediate goods, and our notion of viability isolates the logically prior question that this network-economics literature takes as given---namely, the structural condition (coherence, equivalently the non-singularity of the system's matrix representation) under which a strictly positive viable price system exists at all. This is the precise sense in which our framework bridges Leontief-Sraffian production theory with modern network economics. 

\citet{Baqaee2018} demonstrates how entry and exit in production networks create cascading failures and amplify shocks, challenging the notion that sales shares adequately capture systemic importance. Unlike the relevant notions of centrality in competitive models, systemic importance depends on an industry's role as both supplier and consumer of inputs, as well as market structure. 

\citet{Baqaee2019} extend this analysis, showing that microeconomic details of production structure---network linkages, elasticities of substitution, returns to scale---shape second-order terms in aggregate responses to shocks, going beyond Hulten's theorem. Their work on the microeconomic foundations of aggregate production functions \citep{BaqaeeFarhi2019} establishes that aggregation from heterogeneous micro-level technologies to macro production functions depends critically on the economy's network structure and elasticities of substitution. These contributions, along with their analysis of productivity and misallocation in general equilibrium \citep{BaqaeeFarhi2020}, demonstrate that production network structure fundamentally affects the mapping from microeconomic shocks to aggregate outcomes across different frequencies. 

\citet{AcemogluAzar2020} endogenise the production network itself, allowing each product to be produced with an endogenously chosen subset of other products as inputs, and show that improvements in technology spread through the input-output linkages, reducing all prices and resulting in a denser production network.

The role of production networks in shaping macroeconomic dynamics extends to monetary policy. \citet{HuangLiu2001} analyse a multi-stage production economy, demonstrating that the input-output structure creates strategic complementarities in price-setting. Building on this insight, \citet{LaOTahbazSalehi2022} characterise optimal monetary policy in production networks with nominal rigidities, showing that the optimal policy stabilises a price index with greater weights on larger, stickier, and more upstream industries. Their analysis reveals that network structure fundamentally shapes the transmission of monetary policy and the trade-offs faced by monetary authorities. 

\citet{PelletTahbazSalehi2023} further explore how quantity rigidities and informational frictions in production networks dampen the impact of productivity shocks while amplifying the effects of demand shocks, with the magnitude depending on the network's structure.

\citet{Levine2012} emphasises the ``weakest link'' property where long chains permit specialisation but create vulnerability to failure. This motivates our focus on viability: if any producer fails, downstream producers may also fail, creating cascading disruptions. \citet{Bosker2014} examine how production networks affect economic geography, showing that network structure interacts with trade costs to determine spatial patterns of production.

\paragraph{Bargaining and intermediation}

The literature on bargaining in networks provides important insights for understanding price formation for intermediate goods. \citet{Siedlarek2025} develops a stochastic bargaining model for networked markets with intermediaries, showing that non-essential players---those who can be circumvented through alternative routes---receive zero payoffs as bargaining frictions vanish, while essential players extract positive rents. This result provides microfoundations for thinking about how network position affects bargaining power and price formation in production chains.

\citet{Alviarez2026} develop and estimate a structural model of bargaining in firm-to-firm trade in which both buyers and suppliers exercise market power. In their framework, bilateral markups and the pass-through of cost shocks are determined by two sufficient statistics---the supplier's share in the buyer's purchases and the buyer's share in the supplier's output---and their estimates document substantial buyer bargaining power. This demonstrates how bilateral bargaining shapes the division of surplus and the transmission of cost changes in individual supply relationships. 

\citet{Bizzarri2023} provides a model of general equilibrium oligopoly in input-output networks where all firms have market power on both input and output markets and are fully strategic about their position in the supply chain.

\paragraph{Frictions and trading networks}

\citet{Fleiner2019,Fleiner2022} develop a theory of trading networks with distortionary frictions that make utility imperfectly transferable between agents. They establish existence of competitive equilibria under these conditions and provide cooperative foundations for competitive equilibrium in finite markets with frictions. Their work demonstrates that many structural results---including lattice properties and versions of the rural hospitals theorem \citep{RothSotomayor1990}---extend to environments with trading frictions. While their focus is on exchange networks rather than production networks, their results on how frictions affect equilibrium outcomes inform our analysis of price formation under institutional constraints. \citet{Huremovic2020} analyse how financial shocks propagate through production networks, showing that the interaction between financial frictions and production linkages amplifies aggregate volatility.

The supply chain management literature \citep{Cachon2003} and work on vertical integration \citep{Williamson1985} document how relationship-specific investments and contractual incompleteness shape production organisation. Game-theoretic approaches \citep{KrantonMinehart2001} explore how bilateral relationships affect economic outcomes. Our framework extends this literature by embedding bargaining within a general equilibrium structure.

\subsection{Our contribution}

Our framework distinguishes itself by treating consumption goods and intermediate goods as fundamentally different economic objects requiring distinct pricing mechanisms: competitive markets for the former and bilateral bargaining for the latter. By allowing consumption goods to have competitive prices while intermediate goods have negotiated prices, we capture essential features of real production systems obscured by traditional approaches.

The concept of viability introduced here provides a minimal requirement for economic sustainability, weaker than full equilibrium but essential for any stable configuration. Unlike existence results in \citet{Baqaee2018} focusing on free entry equilibria with external economies of scale, or the general equilibrium frameworks of \citet{BaqaeeFarhi2020} emphasising misallocation and productivity, our viability analysis asks the more fundamental question: which production network structures admit price systems that allow all producers to cover their costs? This is a necessary condition for any equilibrium, whether competitive, oligopolistic, or based on bargaining.

Our main theorem provides a complete characterisation of viability in terms of the structural design of the production system: a structured production system is viable if and only if it is coherent---admitting no circular conversion processes that yield no net output---which in turn is equivalent to the non-singularity of its matrix representation. This equivalence ties the economic requirement that all producers earn positive incomes to a purely structural property of the production technology, thereby contributing to the literature on the Hawkins-Simon condition and positive output price systems in input-output analysis. In particular, acyclic production systems---in which the input-output relationships among professions admit no cycles---are always viable, extending insights from \citet{Levine2012} on production chains to our setting with consumption and intermediate goods. Coherence, however, is strictly weaker than acyclicity: cyclic systems are viable precisely when they are coherent.

We subsequently strengthen viability to complete viability---the requirement that viable prices exist for every strictly positive consumption good price vector---which ensures that the production system remains sustainable regardless of how competitive markets price the final consumption goods. We establish that complete viability stands exactly between two input restrictions: it is guaranteed in coherent systems in which no consumption good serves as an input in any production process, and it in turn requires that no consumption good serves as an input in the production of another consumption good. These results suggest that certain production architectures are inherently more stable than others.

These structural results complement the emphasis on network topology in \citet{Baqaee2018} and \citet{Baqaee2019}, but focus on the static question of sustainability rather than dynamic responses to shocks. Our findings have implications for industrial organisation and economic development. By separating the analysis of viability from market clearing and bargaining, we provide foundations for a richer equilibrium theory that can accommodate the insights from \citet{Siedlarek2025} on bargaining power and network position, from \citet{Alviarez2026} on two-sided market power in supply relationships, and from \citet{Fleiner2019} on frictions in trading networks. This modular approach allows us to build incrementally toward a comprehensive theory of equilibrium in structured production systems.

\subsection{Organisation of the paper}

The remainder of this paper develops the analysis of viability in structured production systems. Section 2 introduces the formal framework, defining structured production systems and their key properties. Section 3 analyses viability, establishing existence results and characterising the relationship between structural properties and viable prices. Section 4 examines complete viability and its connection to input restrictions. Section 5 concludes, outlining the broader research program and discussing implications for economic theory and policy.

\section{Structured production systems}

Our aim is to develop a mathematical model that captures a production system that is founded on structured or chained production processes through fully specialised economic agents.  Hence, production processes are fully expressed in all input and output flows related to all individual production units and the accompanying trade between these units. The final outflows from the professional production system are made up of the produced quantities of consumables that are consumed by the population of economic agents in the production system. As such, the model represents a \emph{structured} supply chain production system based on fully specialised production units, incorporating a static social division of labour.

Each production unit or \emph{producer} in a production chain is represented as a single economic decision maker---an economic agent---and each agent in the production system embodies exactly one production unit in the professional production system. All economic agents are producers and assume some ``professional'' function in the production system. Hence, these producers are fully specialised, each assuming exactly one of a finite number of ``professions'', each represented by a fully specialised (fixed) production plan.

We assume throughout that each producer is fully specialised in the production of a \emph{single} output. This output can be \emph{either} a final consumption good, \emph{or} an intermediate input for productive tasks of other producers.  Consequently, in this model production is ordered through supply chains of fully specialised economic agents. 

In these structured production processes, fully specialised producers trade intermediate goods among themselves. Their trade is solely determined by the technical requirements of the production processes, and these goods are never consumed. The prices of intermediate goods are determined through bargaining, making them at best \emph{partially competitive}.

On the other hand, consumption goods are brought to market in the same way as in any market economy. The supply of these goods is confronted with the demand for them in a standard fashion. Consequently, consumption good prices are determined through the interaction of supply and demand in competitive markets.

\subsection{Mathematical preliminaries}


Throughout the paper, we use a number of mathematical concepts that we list here for the convenience of the reader.

We adopt the following conventions for vector notation. Let $x, y \in \mathbb{R}^n$ be two vectors in the standard $n$-dimensional Euclidean space. We denote $x \geqslant y$ if $x_k \geqslant y_k$ for all $k = 1, \ldots, n$. Next, we say that $x > y$ if $x \geqslant y$ and $x \neq y$. Finally, $x \gg y$ indicates that $x_k > y_k$ for all $k = 1, \ldots, n$. We denote by $e^k$ the $k$-th standard Euclidean unit vector in $\mathbb{R}^n$ and by $0\in \mathbb R^n$ the vector of all zeros.  Thus, in particular, a vector $x \in \mathbb R^n$ is non-negative (non-positive) if and only if $x \geqslant 0$ ($x \leqslant 0$); $x \in \mathbb R^n$ is positive (negative) if and only if $x > 0$ ($x < 0$); and, $x \in \mathbb R^n$ is strictly positive (strictly negative) if and only if $x \gg 0$ ($x \ll 0$). We denote by $\mathbb{R}^n_+$, $\mathbb{R}^n_+ \setminus \{0\}$ and $\mathbb{R}^n_{++}$ respectively the set of non-negative, positive and strictly positive vectors of $\mathbb{R}^n$. 

Let $\mathbb{R}^{n \times m}$ denote the set of $n \times m$ real matrices.  Given a matrix $A\in \mathbb R^{n\times m}$, the transpose of matrix $A$ is denoted by $A^\top\in \mathbb R^{m \times n}$; $0$ denotes the matrix all of whose entries are zero and $I$ denotes the identity matrix, i.e., $I = (\delta_{kh})$, where $\delta_{kh}$ is the Kronecker delta, i.e., $\delta_{kk} = 1$ and $\delta_{kh} = 0$ for $k \neq h$. Following standard convention, $x = (x_k)_k \in \mathbb{R}^n$ is understood as a column matrix in $\mathbb{R}^{n \times 1}$ and then $x^\top=(x_1, \ldots, x_n) \in \mathbb{R}^{1 \times n}$ represents a row matrix. Consistently, given $A \in \mathbb{R}^{n \times m}$, we write $A \geqslant 0$ if $a_{kh} \geqslant 0$ for all $k = 1, \ldots, n$ and $h = 1, \ldots, m$; $A > 0$ if $A \geqslant 0$ and $A$ is not the zero matrix; and finally $A \gg 0$ if $a_{kh} > 0$ for all $k = 1, \ldots, n$ and $h = 1, \ldots, m$. 

A square matrix $A=(a_{kh})$ is a $\mathcal{Z}$-matrix or a matrix of class $\mathcal{Z}$ if $a_{kh}\leqslant 0$ for any $k \neq h$. A $\mathcal{Z}$-matrix $A=(a_{kh})$ is of class $\mathcal{Z}^+$ if it has a positive (main) diagonal, i.e., if $a_{kk}>0$ for any $k$ \citep{Giorgi2022}.

A $\mathcal{Z}$-matrix $A$ satisfies the Hawkins-Simon condition \citep{HawkinsSimon1949} if the leading principal minors of $A$ are all positive. \citet{McKenzie1960} extended the notion of matrices with a \emph{dominant diagonal}, which goes back to the work of Hadamard, as follows:

\begin{definition}\label{def:p.q.d.d.}
	A square real matrix $A=(a_{kh})$ of order $n \in \mathbb N$ is said to have a \textbf{row quasi-dominant diagonal} (a \textbf{column  quasi-dominant diagonal}) if there exist positive numbers $d_1, d_2, \ldots, d_n>0$ such that
	\[
	d_k|a_{kk}|> \sum_{h\neq k}d_h|a_{kh}|\quad {\rm for \,all\,}k=1,\ldots, n \quad \left(d_h|a_{hh}|> \sum_{k\neq h}d_k|a_{kh}|\quad {\rm for \,all\,}h=1,\ldots, n \right) .
	\]
\end{definition}

\noindent
Unlike the notion of dominant diagonal, if $A$ has a row quasi-dominant diagonal, then it has a column quasi-dominant diagonal and vice versa. Therefore, we simply refer to these matrices as having a \emph{quasi-dominant diagonal} (q.d.d.). If, in addition, $a_{kk}>0$ for all $k=1,\ldots, n$, then $A$ is said to have a \emph{positive quasi-dominant diagonal} (p.q.d.d.). \citet{McKenzie1960} shows that:
\begin{numm}
	\item  If $A$ has a q.d.d., then $A$ is nonsingular;
	\item  If $A$ has a p.q.d.d., then all the principal minors of $A$ are positive.
\end{numm}
Theorems 1 and 4 of \citet{Giorgi2023} summarise the state of the literature on investigating equivalent statements of the Hawkins-Simon condition (see also \citet{Berman1994}). In what follows, we list a few equivalent conditions that we use for our analysis.
\begin{lemma} \label{lem:Giorgi}
	Let $A= (a_{kh})$ be a square matrix of order $n \in \mathbb N$ that is of class $\mathcal Z^+$. Then the following statements are equivalent:
	\begin{abet}
		\item There exists some $x \in \mathbb R^n \colon x > 0$ and $Ax \gg 0$;
		\item \emph{(Hawkins-Simon condition)} The leading principal minors of $A$ are all positive;
		\item The matrix $A$ has a positive quasi-dominant diagonal (p.q.d.d.).
	\end{abet}
\end{lemma}

\noindent
We refer to \citet{Berman1994} for the following mathematical preliminaries on matrix algebra. 
	\begin{numm}
		\item A square matrix $A$ is said to be \emph{convergent} if $\lim_{m\to+\infty} A^{m}$ exists and is the zero matrix. 
		\item For a square matrix $A$, the scalar $\rho(A) \;=\; \max\bigl\{ \, |\lambda| : \lambda \text{ is an eigenvalue of }A \, \bigr\}$
		is called its \emph{spectral radius}.  Note that $\rho(A)=\rho(A^{\top})$.
	\end{numm}

\begin{lemma} \label{Fact 1} 
Let $A$ be a square matrix. 
\begin{enumerate}
	\item If $A\geqslant 0$ is a non-negative square matrix, then
	\begin{abet}
		\item $\rho(A)$ is an eigenvalue of $A$.
		\item There exists some $x > 0$ such that $x$ is an eigenvector of $A$ corresponding to $\rho(A)$.
		\item $A$ is convergent if and only if $\rho(A)<1$.
		\item $A$ is convergent if and only if $I-A$ is non-singular and
		$
		(I-A)^{-1} \;=\; \sum_{m=0}^{\infty} A^{m}\geqslant 0.
		$
		\item $\alpha\,x \leqslant Ax \text{ with } x > 0$ implies that $\rho(A)\geqslant \alpha$.
		\item $ Ax \leqslant \beta\, x \text{ with } x\gg 0$ implies that $\rho(A)\leqslant\beta$.
	\end{abet}
	\item If $A = \alpha I - B$ with $B\geqslant 0$ and  if $Ax\geqslant 0$ for some $x\gg 0$, then $\rho(B)\leqslant \alpha$.	
\end{enumerate}
\end{lemma}




\noindent
The interior of a set $X \subseteq \mathbb R^n$ is denoted as $\mathrm{int} \, X \subseteq X \subseteq \mathbb R^n$. Furthermore, if a set $X \subseteq \mathbb R^n$ is contained in a lower-dimensional subspace $Y$ of $\mathbb R^n$, then its relative interior, denoted by $\mathrm{ri} \, X \subseteq Y$, is the interior of $X$ relative to the subspace $Y$. 

A \emph{polytope} in $\mathbb R^n$ is a set that is the convex hull of a finite set of vectors or points in $\mathbb R^n$. Alternatively, by the Minkowski-Weyl Theorem, a polytope can be understood as a bounded set that is the intersection of a finite number of closed half-spaces, $H(y,c) = \{ x \in \mathbb R^n \mid x\cdot y \geqslant c \}$ for some $y \in \mathbb R^n \setminus \{ 0 \}$ and $c \in \mathbb R$, see \citet[Proposition 3.1.2]{Convexity2024}---and \citet[Theorem 19.1]{Rockafellar1970} for the more general class of polyhedra. 


\subsection{Static intermediated production through professions}

In this section we develop a mathematical framework that captures the principles laid out in the introduction. Principally, our approach is based on the distinction between final consumption goods and intermediate goods. This distinction, exemplified in using separate representations of these two types of goods, is central in our model. The model introduced here was initially set out by \citet[Chapter 6]{SDL-2}.

\paragraph{Production of consumables and intermediate goods}

There are $\ell_c \in \mathbb N$ consumption goods, which are produced through processes founded on the use of $\ell_p \geqslant 0$ intermediate goods or products. Consumption goods are considered to be final outputs of the production system and brought to market. 

We introduce $L_c = \{ 1, \ldots , \ell_c \}$ as the corresponding set of consumption goods. These goods are traded in $\ell_c$ independent competitive markets. Throughout we assume that $\ell_c \geqslant 1$, which guarantees a meaningful production framework generating at least one consumptive final output.

On the other hand, intermediate goods or inputs are trucked and bartered in bargaining processes between production units in the supply chains. The set of intermediate goods is $L_p = \{1, \ldots, \ell_p\}$ if $\ell_p \in \mathbb N$ and $L_p = \varnothing$ if $\ell_p =0$.\footnote{In particular, the property that $\ell_p = 0$ refers to the case that all consumption goods are home-produced, which is the foundation of a pure exchange economy \citep{ArrowDebreu1954,Debreu1959}. Indeed, in this case, every production plan is actually equivalent to an initial endowment.}

Within this setting the consumption space is determined as the non-negative orthant of the $\ell_c$-dimensional Euclidean space $\mathbb R^{\ell_c}_+$. On the other hand, all production processes take place within the total commodity space $\mathbb R^{\ell}$, where $\ell = \ell_c + \ell_p$ is the number of all goods in the production system. We assume that all goods in $L$, where $L = L_c \cup L_p$, are tradable in the production system, or in the consumption good markets. 

We assume that each good $k \in L$ is produced by individuals who assume the corresponding profession related to producing good $k$. The profession for producing good $k$ is fixed and determined fully by the production technology for that particular good. Therefore, there are exactly $\ell$ professions and $L$ can also be interpreted as the set of professions in the structured production system. 

We assume that the production of each good $k \in L$ is efficient: Every agent who assumes the profession related to good $k$ is assumed to produce exactly according to the same ``professional'' production plan.\footnote{This corresponds to an implementation of the hypothesis of Marshallian competition between producers \citep{Marshall1890}. The standards of the profession are productively efficient in the sense that they generate the highest surplus per produced unit. Any deviation from these standards would result in lower surpluses and reduced generated incomes.} In particular, a fully specialised professional producer of good $k$ has a given output of $Q^k >0$ of that good, which requires the input of a bundle $y^k \in \mathbb R^{\ell}_+$. This is formalised next.

\begin{definition} \label{def:profession}
	For each good $k \in L$, the corresponding \textbf{profession} is defined by a fully specialised production plan with good $k$ being the only net output. This can be represented as
	\begin{equation}
	\zeta (k) = Q^k \, e^k - y^k \in \mathbb R^{\ell}
	\end{equation}
	where $Q^k >0$ is the fixed generated output quantity of good $k$, and $y^k \in \mathbb R^{\ell}_+$ with $y^k_k=0$ is the required net input vector to generate $Q^k$ units of good $k$.\footnote{The requirement that $y^k_k=0$ indeed implies that the vector $y^k \in \mathbb R^{\ell}_+$ is a \emph{net} input vector rather than a gross input vector. }
	\\
	The mapping $\zeta \colon L \to \mathbb R^{\ell}$ with $\zeta (k) = Q^k \, e^k - y^k$ now represents the corresponding complete \textbf{professional production system}.
\end{definition}

\noindent
Definition \ref{def:profession} introduces every profession of a good as unique in terms of that good. Hence, for two professions $k \neq k'$, it holds that $\zeta(k) \neq \zeta(k')$. This is due to the fact that for each profession $k \in L$, good $k$ is its unique net output.  

We note that the net input vector $y^k$ might require positive input quantities of consumables as well as intermediate goods. From this formalisation we identify good $h \in L$ as a \emph{net intermediate input for the production of} $k \in L$ if $y^k_h >0$, where $y^k_h$ is the net amount of good $h$ that is used in the production of $Q^k$ of good $k$. This is regardless of whether the input is an intermediate or consumption good. Definition \ref{def:profession} imposes that $\zeta (k) \in \mathbb R^\ell$ is a vector in which only the $k$-th coordinate has a positive entry, while all other entries are non-positive. In particular, good $k \in L$ is never a net intermediate input for its own production, since $y^k_k=0$. 

If $y^k=0$, we refer to profession $k \in L$ as a \emph{home production} technology, since it requires no net inputs into its production process, i.e., $\zeta (k) = Q^k e^k$.

\paragraph{Matrix representation of $\zeta$}

It is useful to use a matrix representation of a professional production system $\zeta$. This matrix representation lists all professions and their constituting production plans in an $( \ell \times \ell)$ matrix.  

Formally, denote by $Z$ the $(\ell \times \ell)$ matrix whose rows are defined by $\zeta (k)^\top = (Q^k e^k - y^k)^\top$. Hence, $Z = (z_{kh})_{k,h \in L}$ where $z_{kk} = Q^k >0$ for every $k \in L$ and $z_{kh} = -y^k_h \leqslant 0$ for all $k,h \in L$ with $k \neq h$. Thus,
\begin{equation} \label{eq:Z-matrix}
Z = \begin{pmatrix}
\zeta(1)^\top \\
\zeta(2)^\top \\
\vdots  \\
\zeta(\ell)^\top
\end{pmatrix}=
\begin{pmatrix}
Q^1 & -y^1_2 & \cdots & - y^1_\ell \\
-y^2_1 & Q^2 & \cdots & - y^2_\ell \\
\vdots & \vdots & \ddots & \vdots \\
-y^\ell_1 & -y^\ell_2 & \cdots & Q^\ell
\end{pmatrix}
\end{equation}
It is clear from the construction of $Z$ that $Z$ is actually a matrix of class $\mathcal Z^+$.

Throughout we use the convention that all consumption good production plans $k \in L_c$ are listed in the first $\ell_c$ rows of the matrix $Z$. All intermediate good production plans $k \in L_p$ are listed in rows $\ell_c+1, \ldots , \ell$ of the matrix $Z$, noting that $\ell = \ell_c + \ell_p$.

\paragraph{Definition of structured production systems}

Next, we implement a professional production system $\zeta$ within a population of economic agents that assume professions listed in $\zeta$. This creates a structured production system. Let the finite set of agents in the production system be given by $N$. Throughout, we assume that all professions $k \in L$ are fully implemented in a population. This requires that $\#N \geqslant \ell = \ell_c + \ell_p$, ensuring that all goods can be produced by assigning professions in the professional production system $\zeta$ to certain agents in $N$, and that the assignment of agents to professions from $N$ to $L$ is \emph{surjective} as hypothesised in Definition \ref{def:SPS}. Consequently, each introduced profession $k \in L$ is assumed by at least one, but potentially multiple, agents in $N$.\footnote{Our framework could also accommodate cases where an agent does not adopt a profession; however, such an individual would not contribute to the production system, rendering them irrelevant to the current analysis. To formalise this possibility, one could define a subset $N' \subseteq N$ containing only productive agents, such that $\# N' \geqslant \ell$, and define the surjective function $\gamma: N' \to L$. Nevertheless, the presence of non-productive agents in $N \setminus N'$ has no impact on the structured production system itself. Their role will become relevant in subsequent papers where agents are considered as both consumers and producers, aligning with the \lq consumer-producer\rq\ duality established by \citet{SDL2019}. See also \citet{SDL2020-JEBO} and \citet{SDL2019Core} for the analysis of the Core in these economies.  }

The following definition introduces the main concept that underpins the analysis presented here. We understand a structured production system to be based on the production of goods through objective professions. The net outputs of these professions are either traded in a competitive market (for consumption goods) or traded along a binary relationship between producers (for intermediate goods). 

\begin{definition} \label{def:SPS}
	A \textbf{structured production system} with $\ell_c \geqslant 1$ consumables, $\ell_p \geqslant 0$ intermediate goods and $\#N \geqslant \ell = \ell_c + \ell_p$ economic agents is a tuple $\mathbb S = \langle N, \zeta , \gamma \rangle$ where
	\begin{itemize}
		\item $N$ is a finite set of economic agents, represented as fully specialised producers,
		\item $\zeta \colon L \to \mathbb R^{\ell}$ is a professional production system, describing the profession-based production technology in the system as introduced in Definition \ref{def:profession}, and
		\item $\gamma \colon N \to L$ is an assignment of professions to all economic agents in the production system that is surjective, resulting in a partitioning of $N$ into professional (equivalence) classes $\{ N_k \mid k \in L \}$ with $	N_k = \{ i \in N \mid \gamma (i) = k \}$
	\end{itemize}
	such that there exists some $\omega \in \mathbb R^{\ell_c}_{++}$ with
	\begin{equation} \label{eq:StructuredEconomy}
	\sum_{i \in N} \zeta (\gamma (i)) = Z^\top \, n = \binom{\omega}{0} \in \mathbb R^{\ell\times 1}  ,
	\end{equation}
	where $n = (n_1, \ldots , n_\ell )^\top$, with $n_k = \# N_k$ and $\sum_{k\in L} n_k=\#N$, is the vector of professional class sizes. \\
	Condition (\ref{eq:StructuredEconomy}) is also denoted as the \textbf{structured production hypothesis}. 
\end{definition}

\noindent
The structured production hypothesis stated as (\ref{eq:StructuredEconomy}) implies that all intermediate goods generated within a production system are absorbed into overall production processes. Consequently, the net output of the production system is represented by a strictly positive vector of consumption goods, denoted as $\omega \gg 0$. The structured production hypothesis (\ref{eq:StructuredEconomy}), therefore, is a productive efficiency hypothesis. All intermediate inputs are produced in quantities that \emph{exactly} suffice for the production of the consumable goods.

Furthermore, the structured production hypothesis (\ref{eq:StructuredEconomy}) allows for a meaningful confrontation between the net generated quantities of \emph{all} consumption goods and the generated demand from consumers, as the total output of the professional production system in the production system consists solely of consumption goods.  

A structured production system represents production that corresponds to a generalisation of Leontief's input-output framework \citep{Leontief1936,Leontief1941}. Instead of industrial sectors, we use professions assigned to individual producers in the description of production. Furthermore, the input-output framework is modified to distinguish between consumption and intermediate goods explicitly. The system's objective is redefined as the net production of consumption goods only, even though all goods are produced in positive quantities within the system.
%
%
%
%
%

\begin{remark}
	We note that not every professional production system $\zeta$ on a set of goods $L = L_c \cup L_p$ is {implementable}, in the sense of satisfying the structured production hypothesis (\ref{eq:StructuredEconomy}) for some population of economic agents $N$ and assignment function $\gamma \colon N \to L$. Indeed, the restriction of the population vector $n$ to natural numbers is a non-trivial requirement that can fail even in simple cases. For instance, consider a system where two consumption goods $k_1$ and $k_2$ are produced by a single intermediate input $k_3$, with the production matrix:
	\[
	Z =
	\begin{pmatrix}
	Q^1 & 0 & - \sqrt{2} \\
	0 & Q^2 & 0 \\
	0 & -1 & 1
	\end{pmatrix}
	\]
	For this system to be implementable, the net production of $k_3$ must balance, requiring $n_{k_3} - \sqrt{2} n_{k_1} = 0$. Since $\sqrt{2}$ is irrational, this identity cannot hold for any vector $n \in \mathbb{N}^3$ with $n \gg 0$. As the surjectivity of the assignment $\gamma$ hypothesised in Definition \ref{def:SPS} requires that $n_k \geqslant 1$ for every profession $k$, this demonstrates that $\zeta$ is not implementable. Consequently, in view of Definition \ref{def:SPS}, we restrict our analysis to professional production systems that are indeed implementable.
\end{remark}

\paragraph{Comparison with network representation of production systems}

\citet{SDL-2} introduced economies with production networks which are further explored in \citet{GillesPesce-SP-3}. It is worth noting that every production network in the sense of \citet{GillesPesce-SP-3} is a specific implementation of a structured production system as defined above. The positions in a production network are occupied by fully specialised economic agents, each being assigned a profession defined in Definition \ref{def:profession}. The links in a production network are weighted and describe a quantity of a commodity traded between two economic agents. The in-flows and outflows of each position in the production network exactly add up to the production plan representing the profession of the economic agent constituting this position. 

Therefore, each production network implements a well-defined structured production system. On the other hand, the $\mathcal Z$-matrix representation of some professional production system $\zeta$ itself is an adjacency matrix of some weighted network in only very specific circumstances. This refers to the special case in which it is possible that each profession can be assumed by a single ``representative agent'' to form a structured production system satisfying (\ref{eq:StructuredEconomy}).

It should be clear that each production network contains more information than is contained in the corresponding structured production system. For details we refer to \citet[Chapter 6]{SDL-2} and \citet{GillesPesce-SP-3}.

\subsection{Coherence of production systems}

In the context of a structured production system $\mathbb S$ we now introduce a  notion of coherence that imposes that there are no meaningless cyclic conversion processes of intermediate inputs only. This is formalised as follows. 

\begin{definition} \label{def:Coherence}
	Let $\mathbb S = \langle N, \zeta , \gamma \rangle$ be some structured production system and let $n = (n_1, \ldots , n_\ell )^\top$ be the corresponding vector of professional class sizes. Let $\zeta$ be represented by the matrix $Z$.  \\
	A \textbf{conversion cycle} in $\mathbb S$ is a  vector $\alpha \in \mathbb R^\ell_+ \setminus \{ 0 \}$ with $0 < \alpha \leqslant n$ such that
	\begin{equation}
	Z^\top \alpha = \sum_{k \in L} \alpha_k \, \zeta (k) =0 \in \mathbb{R}^{\ell \times 1}.
	\end{equation}
	The structured production system $\mathbb S$ is \textbf{coherent} (C) if it does not admit any conversion cycles.
\end{definition}

\noindent
Coherence implies that there are no meaningless conversion processes or commodity flows in the production system. This mainly refers to the conversion of intermediate goods into other intermediate goods without these intermediate goods having a destiny in the production of consumption goods as the net output from the production system. 

In the context of the structured production hypothesis ($\ref{eq:StructuredEconomy}$), which implies that the social division of labour results in a \emph{productive} economy that fully exhausts all intermediate goods while yielding a positive surplus of consumables, coherence is the requirement that this division of labour be \emph{strictly purposeful}. Specifically, the production of intermediate goods must be inextricably linked to the eventual creation of consumables. Thus in particular, in a coherent system, no subset of the specialised agents can be engaged in a self-serving production chain that fails to contribute to the economy's final consumption goals.

\begin{remark} \label{rem:Coherence}
	Coherence, as defined in Definition \ref{def:Coherence}, seems weaker than the property that the professional production system $\zeta$ on $L$ has a non-singular matrix representation $Z$ with $\det Z \neq 0$. Indeed, this latter property signifies the linear independence of all professions $\zeta(k)$ for $k \in L$, whereas the absence of conversion cycles leads to a specific form of linear independence based solely on non-negative multipliers. Nevertheless, Theorem \ref{thm:ViabilityStructural} shows that a structured production system $\mathbb{S}=\langle N, \zeta , \gamma \rangle$ is coherent if and only if the matrix $Z$ representing $\zeta$ is non-singular, i.e., $\det Z\neq 0$.
\end{remark}

\noindent The next example illustrates Definition \ref{def:Coherence} and Remark \ref{rem:Coherence} with a structured production system that has an obvious conversion cycle.

\begin{example} \label{ex:Coherence}
	Consider the production system $\mathbb{S}$ with one consumable good $k_1$ and three intermediate goods $k_2$, $k_3$ and $k_4$. The production of $k_1$ only depends on intermediate input $k_4$, while $k_2$ and $k_3$ are converted into each other. A corresponding professional production system $\zeta$ can be represented as
	\[
	Z=\left[ \begin{array}{cccc}
	Q^1&0&0&-Q^4\\
	0&Q^2&-Q^3&0\\
	0&-Q^2&Q^3&0\\
	0&0&0&Q^4	 
	\end{array}\right] \qquad \mbox{ with } Q^i>0\,\mbox{for} \, i=1,2,3,4. 
	\]
	Note that $Z$ has a $2 \times 2$ block that represents that $Q^2$ units of $k_2$ are converted into $Q^3$ units of $k_3$, and vice versa. This indicates that this structured production system is \emph{not} coherent. Indeed, let $\alpha = (0,1,1,0)^\top$, then $Z^\top \alpha = 0$, indicating that $\mathbb S$ contains a conversion cycle. 
	\\[1ex]
	In particular, we compute that
	\[
	\det Z = Q^1 \cdot Q^4 \cdot \det \left[
	\begin{array}{cc}
	Q^2 & - Q^3 \\
	- Q^2 & Q^3
	\end{array}
	\right] =0.
	\]
	This confirms by Remark \ref{rem:Coherence} that $\mathbb S$ is not coherent. 
\end{example}

\section{The viability of structured production systems}\label{sec:Viability}

Next, we investigate how the generated wealth is distributed throughout the prevailing structured production system. We base ourselves on the established principle that wealth distribution is effectuated through the pricing of goods.

Consider a structured production system $\mathbb S = \langle N, \zeta , \gamma \rangle$.  Now, a \emph{price system} in $\mathbb S$ is denoted as an $\ell$-dimensional vector $(p,q) \in \mathbb R^{\ell}_+$. Here $p \in \mathbb R^{\ell_c}_+ \setminus \{ 0 \}$ represents the consumption good prices that emerge in the competitive markets on which these goods are traded. On the other hand, $q \in \mathbb R^{\ell_p}_+$ is the vector of intermediate good prices that emerge in the trade relationships of intermediate good producers and users. 

We have two considerations concerning consumption good prices versus intermediate good prices:
\begin{itemize}
	\item Consumption goods are traded on competitive markets in which supply is confronted with demand for these goods. Therefore, consumption good prices are subject to Walras' Law \citep{Walras1954}, which states that if $m-1$ competitive markets are in equilibrium, the $m$-th market is also balanced. This allows us to normalise the consumption good price vector $p$ such that
	\[
		p \in \bar S = \left\{ (p_1, \ldots , p_{\ell_c} ) \, \left| \, p_k \geqslant 0 \mbox{ for all } k \in L_c \mbox{ and } \sum_{k \in L_c} p_k =1 \right. \right\} .
	\]
	The consumption good price space $\bar S$ is the unit simplex in $\mathbb R^{\ell_c}$ and consists of positive normalised consumption good price vectors $p > 0$ only.
	\item On the other hand, as stated, intermediate good prices are outcomes of bargaining processes between traders of these intermediate goods. These goods are \emph{not} subject to market equilibration processes. Therefore, it is natural to assume that the intermediate good price vector is any non-negative vector $q \in \mathbb R^{\ell_p}_+$. As such, the intermediate goods price vectors $q$ are \emph{not} normalised, but can attain any non-negative value, including zero.
\end{itemize}
Consequently, the \emph{price space} for the structured production system $\mathbb S$ is given by $\bar S \times \mathbb R^{\ell_p}_+$.

\paragraph{Wealth distributions}

A formal representation of a wealth distribution through price adjustment can be formulated as follows. Consider a structured production system $\mathbb S = \langle N, \zeta , \gamma \rangle$ and some price system $(p,q) \in \bar S \times \mathbb R^{\ell_p}_+$. Then for each profession $k \in L$ the \emph{(generated) income} of a professional of type $k$ is determined to be
\[
I_k (p,q) = (p,q) \cdot \zeta (k) = \left\{
\begin{array}{ll}
p_k \, Q^k - (p,q) \cdot y^k \qquad & \mbox{if } k \in L_c \\
q_k \, Q^k - (p,q) \cdot y^k & \mbox{if } k \in L_p
\end{array}
\right.
\]
Now $\{ I_k (p,q) \mid k \in L \}$ represents the wealth distribution at price system $(p,q)$. Critically, in any structured production system, price systems have to be such that all generated incomes are non-negative. This is referred to as the \emph{weak viability} of the price system. Similarly, \emph{viability} of a price system requires that in the generated wealth distribution every profession $k \in L$ receives a strictly positive income.

\subsection{Viable price systems}\label{sec: viable}

In this section, we study the architecture of structured production systems that guarantees our two main system-viability notions, viability (V) and weak viability (WV). The viability property (V) is compared to the structural design properties, acyclicity (A) and coherence (C). From a technical perspective, viability is also equivalent to the Hawkins-Simon condition (HS) and other equivalent conditions found in the literature, in particular those stated in Lemma \ref{lem:Giorgi}. 
 
The following diagram summarises the results that will be established in this section. Here (a) refers to the support of the depicted relationship through Proposition \ref{prop: acyclic implies viable} and Example \ref{ex: viable not acyclic}. Similarly, (b) refers to Remark \ref{rem: viable vs coherent} and (c) refers to Remark \ref{remark:Viable}.
 
\bigskip
\begin{figure}[h]
\centering
\begin{tikzpicture}[
    box/.style       = {draw, minimum size=1cm, font=\large, inner sep=2pt},
    bluebox/.style   = {box, draw=blue, text=blue},
    blackbox/.style  = {box, draw=black},
    impl/.style      = {-{Implies[]}, double, double distance=2pt, line width=0.4pt},
    biimpl/.style    = {blue, {Implies[]}-{Implies[]}, double, double distance=2pt, line width=0.4pt},
    redno/.style     = {red, -{Implies[]}, double, double distance=2pt, line width=0.4pt},
    lab/.style       = {blue, font=\large},
]

\node[blackbox]  (Atop) at (0,3)      {$A$};
\node[bluebox]  (V)    at (0,0)      {$\mathbf V$};
\node[bluebox]  (HS)   at (-3.4,0)   {$HS$};
\node[bluebox] (Amid) at (2.7,0)    {$\mathbf C$};
\node[blue]     (detZ) at (5.6,0)    {$\det Z \neq 0$};
\node[blackbox] (wV)   at (0,-3)     {$WV$};

\begin{scope}[on background layer]
  \node[draw=violet, line width=1.6pt, rounded corners=10pt,
        fit=(V)(Amid)(detZ), inner sep=11pt] (thm) {};
\end{scope}
\node[violet, font=\large, anchor=south east]
      at ([xshift=-6pt, yshift=3pt]thm.north east) {Theorem \ref{thm:ViabilityStructural}};

\draw[biimpl] (HS.east)  -- (thm.west);
\draw[biimpl] (V.east)   -- (Amid.west);
\draw[biimpl] (Amid.east)-- (detZ.west);
\node[lab] at ($(HS)!0.5!(thm.west) + (0,0.55)$) {(c)};

\draw[impl] ([xshift=-7pt]Atop.south) -- ([xshift=-7pt]V.north);
\draw[redno] ([xshift=10pt]V.north) -- ([xshift=10pt]Atop.south);
\draw[red, line width=1.2pt]
      ($(V.north)!0.5!(Atop.south) + (-2pt,-6pt)$) -- ++(24pt,12pt);
\draw[red, line width=1.2pt]
      ($(V.north)!0.5!(Atop.south) + (-2pt,6pt)$)  -- ++(24pt,-12pt);
\node[lab] at ($(V.north)!0.5!(Atop.south) + (1.4,0)$) {(a)};

\draw[impl] ([xshift=-7pt]V.south) -- ([xshift=-7pt]wV.north);
\draw[redno] ([xshift=10pt]wV.north) -- ([xshift=10pt]V.south);
\draw[red, line width=1.2pt]
      ($(wV.north)!0.5!(V.south) + (-2pt,-6pt)$) -- ++(24pt,12pt);
\draw[red, line width=1.2pt]
      ($(wV.north)!0.5!(V.south) + (-2pt,6pt)$)  -- ++(24pt,-12pt);
\node[lab] at ($(wV.north)!0.5!(V.south) + (1.4,0)$) {(b)};

\end{tikzpicture}
\caption{Schematic representation of conceptual relationships in Section 3.1} \label{fig:relations3}
\end{figure}

\noindent
The formal analysis of viability of a structured production system is captured by the following definition of our two main conceptions, viability and weak viability.

\begin{definition}
	Let $\mathbb S = \langle N, \zeta , \gamma \rangle$ be a structured production system. 
	\begin{abet}
		\item The structured production system $\mathbb S$ is \textbf{weakly viable} (WV) if 
		\begin{equation*}  
		\Delta' = \left\{ \left. (p,q) \in \bar S \times \mathbb R^{\ell_p}_+ \, \right| \, I_k (p,q) \geqslant 0 \mbox{ for all } k \in L \, \right\} \neq \varnothing
		\end{equation*}
		\item The structured production system $\mathbb S$ is \textbf{viable} (V) if 
		\begin{equation*}  
		\Delta = \left\{ \left. (p,q) \in \bar S \times \mathbb R^{\ell_p}_+ \, \right| \, I_k (p,q) > 0 \mbox{ for all } k \in L \, \right\} \neq \varnothing
		\end{equation*}
		We refer to any price system $(p,q) \in \Delta$ as a \textbf{viable price system} in $\mathbb S$.
	\end{abet}
\end{definition}

\noindent
Viability ensures that every producer earns sufficient income to justify continued participation, thereby preventing the systemic supply chain failures that would arise if any producer in the production chain were to cease operations.

\begin{remark}\label{rem: viable vs coherent}
	It is clear that viability implies weak viability, i.e., $\Delta\neq \varnothing \Rightarrow \Delta'\neq \varnothing$. The converse need not be true. Indeed, the structured production system considered in Example \ref{ex:Coherence} is not viable because $I_{k_2} (p,q_1,q_2,q_3) = q_1Q^2 -q_2Q^3 >0$ and $I_{k_3} (p,q_1,q_2,q_3) = -q_1Q^2 +q_2Q^3 >0$ are in direct contradiction. However, this structured production system is weakly viable for $\left(1, \frac{q_2Q^3}{Q^2}, q_2,q_3\right)$ where $q_2\in \mathbb{R}_+$ and $0 \leqslant q_3\leqslant \frac{Q^1}{Q^4}$, meaning that weak viability does not imply viability. 
\end{remark}

\noindent
From a mathematical point of view, viability is closely related to the existence of a Leontief price vector in classical input-output analysis of production systems initiated by \citet{Leontief1936,Leontief1941}. This is explored in the next remark. 

\begin{remark} \label{remark:Viable}
	The definition of viability links to the theory of the Hawkins-Simon condition (HS) through Lemma \ref{lem:Giorgi}. Indeed, we claim that:
	\begin{numm}
		\item The set of viable price systems $\Delta$ consists of strictly positive price vectors only, implying that
		\[
		\Delta \subseteq \left( \bar S \cap \mathbb R^{\ell_c}_{++} \right) \times \mathbb R^{\ell_p}_{++} .
		\]
		\item $\Delta \neq \varnothing$ if and only if there exists some $(p,q) \in \mathbb R^\ell$ with $(p,q) >0$ such that $Z (p,q) \gg 0$ if and only if any of the properties listed in Lemma \ref{lem:Giorgi} are valid.
	\end{numm}
	To show claim (i), note in particular that, if there exists some $k\in L_c$ such that $p_k=0$, then
	\[
	I_k(p,q)=-\sum_{h\in L_c \setminus \{ k \}}p_h y_h^k - \sum_{h \in L_p} q_h y^k_h \leqslant 0
	\]
	which is a contradiction to viability. Hence, $p\gg 0$. \\ Similarly, if $q_k= 0$ for some $k \in L_p$, then  $I_k(p,q) \leqslant 0,$ which is absurd. Therefore, $p \gg 0$ as well as $q \gg 0$, showing (i). \\[1ex]
	Note that if $(p,q)\in \Delta$, then $Z  (p,q)\gg0$. Hence, to show claim (ii), we only have to show that any of the properties listed in Lemma \ref{lem:Giorgi}  implies viability. Assume, in particular, that there exists some $(p,q) \in \mathbb R^\ell$ with $(p,q) >0$ such that $Z (p,q) \gg 0$. A similar argument as used to show (i) leads to the conclusion that $Z (p,q) \gg 0$ implies that $(p,q) \gg 0$. Now, define $(\tilde{p},\tilde{q})$ as $\tilde{p}:=\tfrac{p}{\sum_{k\in L_c}p_k} $ and $\tilde{q}:=\tfrac{q}{\sum_{k\in L_c}p_k} $.
	\\
	Then $\tilde{p}\in \left( \bar{S} \cap \mathbb R^{\ell_c}_{++} \right)$ and $\tilde{q}\in \mathbb{R}_{++}^{\ell_p}$.  By the linearity of the inner product we have that $I_k(\tilde{p},\tilde{q})>0$ for all $k \in L$, that is $(\tilde{p}$,$\tilde{q})\in \Delta$. Hence, $\Delta \neq \varnothing$. This shows claim (ii).
\end{remark}

\noindent
Remark \ref{remark:Viable} is illustrated in the next example.

\begin{example} \label{ex:CompleteViable}
	We construct a specific structured production system $\mathbb S$ based on three economic agents and three goods, two consumption goods $k_1$ and $k_2$ ($\ell_c =2$) and one intermediate input $k_3$ ($\ell_p =1$).
	\\
	The agent set is given by $N = \{1,2,3 \}$, of which exactly one agent is assigned to each of the three corresponding professions, e.g., $\gamma(1)=k_1, \gamma(2)=k_2$ and $\gamma(3)=k_3$. The professional production system $\zeta$ for these three producers is introduced in its $(3 \times 3)$ $\mathcal Z$-matrix representation as defined in (\ref{eq:Z-matrix}):
	\[
	Z =
	\begin{pmatrix}
	Q^1 & 0 & - \alpha Q^1 \\
	0 & Q^2 & - \beta Q^2 \\
	0 & 0 & \alpha Q^1 + \beta Q^2
	\end{pmatrix}
	\qquad \mbox{where } Q^1,Q^2 >0 \mbox{ and } \alpha , \beta \in[0,1] \mbox{ with }  (\alpha , \beta) \neq (0,0).\footnotemark
	\]
\footnotetext{The property that $(\alpha , \beta) \neq (0,0)$ is due to the hypothesis that the producer of the intermediate good is assumed to generate a non-negligible output, $Q^3=\alpha Q^1+\beta Q^2 >0$. This is based on the definition of a production plan in Definition \ref{def:profession}.}
	Clearly the two consumption good producers $k_1$ and $k_2$ also generate a positive output of $Q^1>0$ and $Q^2>0$, respectively, using a fractional input of the intermediate input of $0 \leqslant \alpha \leqslant 1$ and $0 \leqslant \beta \leqslant 1$ per unit of output, respectively. 
	\\
	Note also that $\mathbb S$ is coherent, since $\det Z = Q^1Q^2 ( \alpha Q^1 + \beta Q^2) >0$ for the indicated values of $Q^1,Q^2, \alpha$ and $\beta$. 
	
	\medskip\noindent
	Next consider a price system with $p \in [0,1]$ referring to the competitive price of $k_1$, implying that under normalisation the competitive price of $k_2$ is $1-p$. Moreover, we let $q \geqslant 0$ denote the bargaining exchange rate of the single intermediate input. A price system can now be represented by $(p,q) \in [0,1] \times \mathbb R_+$.\footnote{This is a slight abuse of notation that is used throughout the paper, but simplifies the representation. Formally, the price system needs to be represented as $(p,1-p , q)$.}
	\\
	From this we compute the corresponding incomes as follows
	\begin{align*}
		I_{k_1} (p,q) & = p \, Q^1 - \alpha Q^1 \, q = (p- \alpha q) \, Q^1 \\
	I_{k_2} (p,q) & = (1-p) \, Q^2 - \beta Q^2  \, q = (1-p - \beta q ) \, Q^2\\
	I_{k_3} (p,q) & = (\alpha Q^1 + \beta Q^2) \, q 
	\end{align*}
	The set of viable price systems $(p,q)$ is now determined by the three inequalities $I_{k_i} (p,q) > 0$ for $i=1,2,3$. This results into
	\[
	\Delta = \left\{ \left. (p,q) \in (0,1) \times \left( 0 , \tfrac{1}{\alpha + \beta} \right) \, \right| \, \alpha q < p < 1 - \beta q \, \right\}
	\]
	Hence, the production system $\mathbb S$ is viable in the sense that $\Delta \neq \varnothing$ for all admissible parameter values $\alpha , \beta \in [0,1]$ with $(\alpha , \beta ) \neq (0,0)$. This is graphically depicted in Figure \ref{fig:ExViable} below, which projects $\Delta$ on the price space $[0,1] \times \mathbb R_+$ for the case that $0< \beta < \alpha <1$.
\end{example}

\begin{figure}[h]
	\centering
	\begin{tikzpicture}[scale=1.5]
	\draw[-latex] (0,0) -- (7,0) node[right] {$q$};
	\draw[-latex] (0,0) -- (0,4) node[above] {$p$};
	
	\node[below left] at (0,0) {$0$};
	
	\draw (0,3) -- (6,3) node[pos=0, left] {$1$};
	
	\fill[lightgray] (0,0) -- (0,3) -- (30/11,18/11) -- (0,0);
	
	\draw[red, thick] (0,0) -- (5,3) node[pos=1.02, above right] {$p=\alpha q$};
	\draw[red, dashed] (5,3) -- (5,0) node[below] {$\frac{1}{\alpha}$};
	\fill[red] (5,3) circle (1pt);
	
	\draw[darkred, dashed] (2.725,1.6) -- (2.725,0) node[below] {$\frac{1}{\alpha + \beta}$};
	
	\draw[blue, thick] (0,3) -- (6,0) node[pos=1, below] {$\frac{1}{\beta}$};
	\fill[blue] (6,0) circle (1pt);
	\node[blue] at (6,0.5) {$p = 1 - \beta q$};
	
	\node at (1.25,1.5) {$\Delta$};
	\end{tikzpicture}
	\caption{Viable price systems discussed in Example \ref{ex:CompleteViable}} \label{fig:ExViable}
\end{figure}

\subsection{Structural analysis of viability}\label{sec: Structural analysis of viability}
The following main theorem establishes fundamental connections between viability and the system's structural coherence.

\begin{theorem} \label{thm:ViabilityStructural}
	Let $\mathbb S = \langle N, \zeta , \gamma \rangle$ be a structured production system. Then the following statements are equivalent:
	\begin{numm}
		\item The professional production system $\zeta$ has a matrix representation $Z$ with $\det Z \neq 0$.
		\item $\mathbb S$ is coherent.
		\item $\mathbb S$ is viable.
	\end{numm}
\end{theorem}

\noindent
A proof of Theorem \ref{thm:ViabilityStructural} is provided in Appendix \ref{sec: app A1} of this paper.

\begin{remark} \label{rem:Genericity}
	In the space of real square matrices $\mathbb R^{\ell \times \ell}$ the set of regular matrices $\{ M \in \mathbb R^{\ell \times \ell} \mid \det M \neq 0 \}$ constitutes a negligible set of Lebesgue measure 0. Consequently, from the equivalence of coherence and non-singularity of $Z$, the coherence property on the class of structured production systems is generic. 
\end{remark}

\paragraph{Some topological properties of $\Delta$ and $\Delta'$}

We investigate some properties of the sets of viable and weakly viable price systems. These properties are stated in the following proposition for coherent structured production systems only.

\begin{proposition} \label{prop:DeltaProps}
	Let $\, \mathbb S = \langle N, \zeta , \gamma \rangle$ be a coherent structured production system. Then, 
	\begin{abet}
		\item $\Delta' \subseteq  \bar S \times \mathbb R^{\ell_p}_+$ is a (compact) polytope in $\mathbb R^{\ell}$. 
		\item  $\Delta'$ is the closure of $\Delta$, i.e., $\Delta' = \mathrm{cl} \, \Delta$, and the relative interior of $\Delta'$ is the relative interior of $\Delta$, i.e., $\mathrm{ri} \, \Delta' = \mathrm{ri} \, \Delta$. 
	\end{abet}
\end{proposition}

\noindent
A proof of Proposition \ref{prop:DeltaProps} is provided in Appendix \ref{sec: app A2} of this paper.

\medskip
\noindent Note that Example \ref{ex:Coherence} shows that the coherence assumption in Proposition \ref{prop:DeltaProps} cannot be dispensed with. In particular, it demonstrates that if $\mathbb S$ is not coherent, it is possible that $\mathrm{ri} \, \Delta' \neq \mathrm{ri} \, \Delta$. Indeed, since $\Delta'$ is a non-empty convex set, Theorem 6.2 of \citet{Rockafellar1970} implies that $\mathrm{ri} \, \Delta'$ is a convex set with the same dimension as $\Delta'$. Consequently, $\mathrm{ri} \, \Delta' \neq \varnothing$, whereas in this case $\Delta = \varnothing$ and thus $\mathrm{ri} \, \Delta = \varnothing$.

\paragraph{Acyclic production systems}

We formalise next how the professional production system $\zeta$ on $\mathbb R^\ell$ exhibits the property that all production is meaningful and results in the targeted, effective production of the final consumption goods that drive wealth generation in such a system. We then prove that this condition, termed acyclicity, is sufficient for the existence of viable prices. This formalisation captures the idea that production processes are purposeful and lead most effectively to the production of the resulting consumptive outputs. 

\begin{definition} \label{def:Acyclic}
	Let $\zeta \colon L \to \mathbb R^\ell$ be a professional production system.
	\begin{abet}
		\item A \textbf{cycle} in the professional production system $\zeta$ is an ordered subset $\{ k_1, \ldots , k_m \} \subseteq L$ with $m \in \mathbb N \setminus \{ 1 \}$ such that
		\begin{numm}
			\item $k_m$ is a net intermediate  input for the production of $k_1$, i.e., $y^{k_1}_{k_m} >0$, and
			\item for every $i =1, \ldots ,m-1$ it holds that $k_{i}$ is a net intermediate  input for the production of $k_{i+1}$, i.e., $y_{k_i}^{k_{i+1}} >0$.
		\end{numm} 
		\item The professional production system $\zeta$ is \textbf{acyclic} if it does not admit any cycles.
	\end{abet}
	A structured production system $\mathbb S = \langle N, \zeta , \gamma \rangle$ is said to be \textbf{acyclic} (A) if so is the professional production system $\zeta$.
\end{definition}

\noindent
Acyclicity of a professional production system $\zeta$ is a rather strong consistency and efficiency property. It excludes any cycle to be possible in any implementation of a production system by assigning economic agents to the professions represented in $\zeta$. 

It is straightforward to establish the next reformulation of the acyclicity property. It rephrases the absence of productive input cycles by stipulating that all multiplications of cross-inputs are zero.

\begin{remark} \label{rem:Acyclicity}
	The professional production system $\zeta \colon L \to \mathbb R^\ell$ is acyclic if and only if for every ordered subset $\{ k_1, \ldots , k_m \} \subseteq L$ with $m \in \mathbb N \setminus \{ 1 \}$ it holds that
	\begin{equation*}
	y^{k_1}_{k_m} \cdot \prod_{i =1}^{m-1} \, y_{k_i}^{k_{i+1}} =0.
	\end{equation*}
\end{remark}

\noindent
The definition and meaning of acyclicity is illustrated in the next example. 

\begin{example} \label{ex:Acyclicity}
	Consider a professional production system $\zeta$ with two consumables, $k_1$ and $k_2$, and one intermediate good $k_3$. Consumable $k_1$ is produced with two units of the intermediate input $k_3$, represented by the production plan $\zeta(k_1)= (Q^1, 0 ,-2)^\top $, where $Q^1>3$. Consumable $k_2$ is produced based on three units of the consumable $k_1$, resulting in the corresponding production plan $\zeta (k_2) = (-3,Q^2,0)^\top $ with $Q^2>1$. Finally, the intermediate input $k_3$ is produced through the production plan $\zeta (k_3)= (0,-1,2)^\top $, using $k_2$ as an input. \\
	The $(3 \times 3)$ matrix representation of $\zeta$ as defined by (\ref{eq:Z-matrix}) can be given as
	\[
	Z =
	\begin{pmatrix}
	Q^1 & 0 & - 2 \\
	-3 & Q^2 & 0 \\
	0 & -1 & 2
	\end{pmatrix}
	\]
	We identify that there is actually a productive cycle in this simple professional production system. Indeed, the ordered set $\{ k_1, k_2,k_3 \}$ defines a cycle with $y^{k_2}_{k_1} =3 >0$, $y^{k_3}_{k_2} =1 >0$ and $y^{k_1}_{k_3} =2 >0$. Hence, $  y^{k_2}_{k_1}\cdot y^{k_3}_{k_2}\cdot y^{k_1}_{k_3}  = 3 \cdot 1 \cdot 2 =6 \neq 0$ shows that there is indeed a complete production cycle in this professional production system (see Remark \ref{rem:Acyclicity}).  
\end{example}

\noindent
We now show that acyclicity ensures the existence of viable prices. The proof, provided in Appendix \ref{sec: app Acyc}, is algorithmic and based on the iterative Gaussian elimination method, which guarantees the absence of cycles at each step and allows one to compute a viable price for any given positive income level. We note that the assertion also follows indirectly: an acyclic professional production system admits a simultaneous reordering of its professions under which $Z$ is triangular with a strictly positive diagonal, implying $\det Z \neq 0$, so that viability follows from Theorem \ref{thm:ViabilityStructural}. However, the algorithmic proof of this proposition adds further understanding and computational elements to the analysis of our framework.

\begin{proposition}\label{prop: acyclic implies viable}
 Let $\mathbb S = \langle N, \zeta , \gamma \rangle$ be an acyclic structured production system. Then  $\mathbb S$ is viable. 
\end{proposition}

\noindent
The next example examines whether the converse of the implication in Proposition \ref{prop: acyclic implies viable} holds. It shows that, in fact, there exist viable structured production systems that are not acyclic.

\begin{example}\label{ex: viable not acyclic}
	Consider a structured production system with three economic agents $N=\{1,2,3\}$ with $\ell_c =2$ and $\ell_p=1$. Agent $3$ produces the intermediate input, while $1$ and $2$ produce separate consumables. Assuming that the output of $1$ is also used as an input for $2$, the resulting system can be constructed to be non-acyclic. \\
	For example, we can select $\zeta$ to be represented by its matrix form with $\zeta (1)$, $\zeta (2)$ and $\zeta (3)$ listed in this order:
	\[
	Z =
	\begin{pmatrix}
	Q^1 & 0 & -Q^3 \\
	-Q^3 & Q^2 & 0 \\
	0 & -Q^3 & Q^3
	\end{pmatrix}
	\qquad \mbox{where } Q^i=Q^3+x_i\, \mbox{for \,} i=1,2 \,\,\mbox{and with } Q^3,x_1,x_2>0.
	\]
	The resulting net output is given by $\zeta (1) + \zeta (2) + \zeta ( 3) =(x_1,x_2,0)^\top$.
	\\
	Noting that $ y^2_1= y^3_2 = y^1_3 =Q^3>0$, we conclude that there is indeed a non-zero cycle in this production system. Hence, by Remark \ref{rem:Acyclicity}, the  structured production system  $\mathbb S$ is not  acyclic. On the other hand, since $\det Z =(Q^3+x_1)(Q^3+x_2)Q^3 - (Q^3)^3 = (x_1x_2 + Q^3x_1 + Q^3x_2)Q^3 >0$, by Theorem \ref{thm:ViabilityStructural} we get that this structured production system $\mathbb S$  is coherent and   viable.  Hence, this shows that viability does not imply acyclicity. 
	
	\noindent For sake of completeness, we now compute the set $\Delta$ and we note that $Z$ satisfies McKenzie's positive quasi-dominant diagonal (p.q.d.d.) property (see Remark \ref{remark:Viable}).
	
	\noindent To this end, let $q \geqslant 0$ be the intermediate good price and $p \in [0,1]$ the price of $1$'s output, making $(1-p)$ the price of $2$'s output. A price system $(p,q)$ is viable if and only if
	\begin{align*}
	I_1 (p,q) & = Q^1p-Q^3q= (Q^3+x_1)p - Q^3q >0 \\
	I_2 (p,q) & = Q^2(1-p) -Q^3p = (Q^3+x_2)(1-p) -Q^3p = (Q^3+x_2) - (2Q^3+x_2)p >0\\
	I_3 (p,q) & = Q^3q- Q^3 (1-p) = Q^3(p+q)-Q^3 >0 
	\end{align*}
	Hence, $\Delta$ is determined by three inequalities:
	\[
	p+q >1 \qquad \mbox{and} \qquad \frac{Q^3}{Q^3+x_1}  q < p < \frac{Q^3+x_2}{2Q^3+x_2}
	\]
	A viable price is, for example, given by $(p^*,q^*)$ with $p^* = \tfrac{1}{2}$ and $q^* = \tfrac{1}{2} + \varepsilon$, where  $0< \varepsilon < \tfrac{x_1}{2Q^3}$. Indeed, obviously $p^*+q^*=1+ \varepsilon >1$ and
	\[
	\frac{Q^3}{Q^3+x_1} q^* < \frac{Q^3}{Q^3+x_1}  \left( \tfrac{1}{2} + \frac{x_1}{2Q^3} \right) = \tfrac{1}{2} = p^* < \frac{Q^3+x_2}{2Q^3+x_2}
	\]
	for all $x_2>0$. 
	\\
	Furthermore, $Z$ satisfies McKenzie's positive quasi-dominant diagonal (p.q.d.d.) property for $d_1=d_2=1$ and $d_3=\frac{2Q^3+x_2}{2Q^3}$, where we verify the column criterion of Definition \ref{def:p.q.d.d.}. Indeed,
	\begin{gather*}
	d_1|z_{11}| =Q^3+x_1>Q^3=d_2 Q^3+d_3 0=\sum_{k\neq 1} d_ky^k_1\\
	d_2|z_{22}| =Q^3+x_2>Q^3+\frac{x_2}{2}=d_10+d_3 Q^3=\sum_{k\neq 2} d_ky^k_2\\
	d_3|z_{33}| =\frac{2Q^3+x_2}{2Q^3}Q^3=Q^3+\frac{x_2}{2}>Q^3=d_1 Q^3+d_2 0=\sum_{k\neq 3} d_ky^k_3.
	\end{gather*}
	This confirms that this structured production system is indeed viable. 
\end{example}

\section{Complete viability of structured production systems}\label{sec: Complete Viability of structured production systems price systems}

 In this section, the notions of viability and weak viability of a structured production system are strengthened as complete viability (CV) and weak complete viability (WCV). We link these properties to the feedback loops of the professional production system $\zeta$ by introducing the concepts of the restricted input condition (RIC) and the weak restricted input condition (WRIC). Figure \ref{fig:relations4} summarises the relationships that will be established in this section between these concepts and the viability (V) and weak viability (WV) properties discussed in the previous section.
 
 In particular, the relationships are supported by the following alphabetically ordered results:
 \begin{abet}
 	\item Proposition \ref{prop:CV-WCV};
 	\item Remark \ref{remark: CV implies V};
 	\item Example \ref{ex:Coherence} and Remark \ref{remark: CV implies V};
 	\item Theorem \ref{thm:CompleteViable}(a) and Example \ref{ex:notRIC};
 	\item Theorem \ref{thm:CompleteViable}(b) and Example \ref{ex:notWRIC};
 	\item Example \ref{ex:notRIC}; and
 	\item Example \ref{ex:notRIC}.
 \end{abet}
	
\bigskip
\begin{figure}[h]
	\centering
	\begin{tikzpicture}[
    box/.style       = {draw, minimum size=1cm, font=\large, inner sep=2pt},
    impl/.style      = {-{Implies[]}, double, double distance=2pt, line width=0.4pt},
    redno/.style     = {red, -{Implies[]}, double, double distance=2pt, line width=0.4pt},
    redbiimpl/.style = { {Implies[]}-{Implies[]}, double, double distance=2pt, line width=0.4pt},
    lab/.style       = {blue, font=\large},
]

\node[box] (V)    at (0,3.2)    {$\mathbf{V}$};
\node[box] (CV)   at (-5.5,1.8) {$\mathbf{CV}$};
\node[box] (WCV)  at (0,0)      {$\mathbf{WCV}$};
\node[box] (WV)   at (5.5,0)    {$\mathbf{WV}$};
\node[box] (RIC)  at (-3,-3.4)  {$RIC$};
\node[box] (WRIC) at (3,-3.4)   {$WRIC$};

\draw[redbiimpl] (CV.east) -- ++(2.0,0);
\node[font=\Large] at (-2.65,1.8) {$\&$};
\draw[decorate, decoration={brace, amplitude=12pt}]
      (-1.95,-0.7) -- (-1.95,3.95);
\node[lab] at (-4.0,2.25) {(a)};

\draw[impl]  ([yshift=7pt]WCV.east) -- ([yshift=7pt]WV.west);
\draw[redno] ([yshift=-7pt]WV.west) -- ([yshift=-7pt]WCV.east);
\rx{($(WCV.east)!0.5!(WV.west)+(0,-0.235)$)}
\node[lab] at ($(WCV.east)!0.5!(WV.west)+(0,0.8)$) {(b)};

\draw[redno] ([xshift=-9pt]V.south)   -- ([xshift=-9pt]WCV.north);
\draw[redno] ([xshift=9pt]WCV.north)  -- ([xshift=9pt]V.south);
\rx{($(V.south)!0.5!(WCV.north)+(-0.32,0)$)}
\rx{($(V.south)!0.5!(WCV.north)+(0.32,0)$)}
\node[lab] at (1.15,1.6) {(c)};

\draw[impl]  ($(RIC.north east)+(-0.14,0.12)$) -- ($(WCV.south west)+(-0.14,0.12)$);
\draw[redno] ($(WCV.south west)+(0.14,-0.12)$) -- ($(RIC.north east)+(0.14,-0.12)$);
\rx{($(RIC.north east)!0.5!(WCV.south west)+(0.14,-0.12)$)}
\node[lab] at (-2.35,-1.5) {(d)};

\draw[impl]  ($(WCV.south east)+(-0.14,-0.12)$) -- ($(WRIC.north west)+(-0.14,-0.12)$);
\draw[redno] ($(WRIC.north west)+(0.14,0.12)$)  -- ($(WCV.south east)+(0.14,0.12)$);
\rx{($(WCV.south east)!0.5!(WRIC.north west)+(0.14,0.12)$)}
\node[lab] at (2.35,-1.5) {(e)};

\draw[impl]  ([yshift=7pt]RIC.east)   -- ([yshift=7pt]WRIC.west);
\draw[redno] ([yshift=-7pt]WRIC.west) -- ([yshift=-7pt]RIC.east);
\rx{($(RIC.east)!0.5!(WRIC.west)+(0,-0.235)$)}
\node[lab] at ($(RIC.east)!0.5!(WRIC.west)+(0,-0.85)$) {(f)};

\draw[redno] (CV.south) -- (RIC.north west);
\rx{($(CV.south)!0.5!(RIC.north west)$)}
\node[lab] at (-5.25,-0.75) {(g)};

\end{tikzpicture}
	\caption{Overview of relationships between complete viability concepts in Section 4} \label{fig:relations4}
\end{figure}

\subsection{Completely viable price systems}\label{sec: compl viable}

The introduced notions of viability of a structured production system can be strengthened to notions that require that \emph{every} consumption price vector $p \in \bar S$ can be supported as viable through the selection of an appropriate intermediate good price system $q \in \mathbb R^{\ell_p}_+$. 

\begin{definition}
	Let $\mathbb S = \langle N, \zeta , \gamma \rangle$ be a structured production system. 
	\begin{abet}
		\item The structured production system $\mathbb S$ is \textbf{weakly completely viable} (WCV) if for every consumption good price vector $p \in \bar S $ there exists at least one intermediate good price vector $q \in \mathbb R^{\ell_p}_+$ such that $(p,q) \in \Delta'$  is weakly viable, i.e., $I_k (p,q) \geqslant 0$ for all $k \in L$.
		\item The structured production system $\mathbb S$ is \textbf{completely viable} (CV) if for every strictly positive consumption good price vector $p \in \bar S \cap \mathbb R^{\ell_c}_{++}$ there exists at least one intermediate good price vector $q \in \mathbb R^{\ell_p}_{++}$ such that $(p,q) \in \Delta$ is viable, i.e., $I_k (p,q) > 0$ for all $k \in L$.
	\end{abet}
\end{definition}

\noindent
As illustrated in Figure \ref{fig:ExViable} for the case $0< \beta < \alpha <1$, the structured production system considered in Example \ref{ex:CompleteViable} satisfies the complete viability property. In particular, for any $0<p<1$ the corresponding viable intermediate input prices are given by $0 < q < \min \left\{ \tfrac{1-p}{\beta} , \tfrac{p}{\alpha} \, \right\}$. This implies that $\mathbb S$ indeed is completely viable.

\begin{remark}\label{remark: CV implies V}
	It is obvious that any (weakly) completely viable structured production system is (weakly) viable. The converse need not be true as the following example shows. \\
	Consider a production system $\mathbb S$ with two consumables denoted as $k_1$ and $k_2$ and one intermediate good $k_3$. We use a matrix representation for the professional production system $\zeta$ given by
	\[
	Z =
	\begin{pmatrix}
	Q^1 & 0 & -1 \\
	-1 & Q^2 & -1 \\
	0 & 0 & 2
	\end{pmatrix}
	\qquad \mbox{where } Q^1=x+1,\,\,\mbox{with}\,\,Q^2,x>0 .
	\]
	Each of these three professions is assumed by a single economic agent. Therefore, the total output of this production system is now given by $1 \cdot \zeta (k_1) + 1 \cdot \zeta (k_2) + 1 \cdot \zeta (k_3) = (x,Q^2,0)^\top$, implying that the total consumptive output of the production system is given by $\omega = (x,Q^2)^\top \gg 0$. \\
	Note that  the structured production system is coherent as $\det Z = 2Q^1Q^2 >0$ and viable (see Theorem \ref{thm:ViabilityStructural}).  \\[1ex]
	Next, let $p \in [0,1]$ represent the competitive market price of $k_1$, which implies that consumable $k_2$ has a competitive market price of $1-p$. Furthermore, we let $q \geqslant 0$ stand for the price of the intermediate input $k_3$. We arrive at the following income equations:
	\begin{align*}
	I_{k_1} (p,q) & = Q^1p -q \\
	I_{k_2} (p,q) & = Q^2 (1-p) -p -q = Q^2-q - (Q^2+1)p\\
	I_{k_3} (p,q) & = 2q 
	\end{align*}
	Imposing viability means that $I_{k_i} (p,q)>0$ for all $i=1,2,3$. The resulting set of viable price systems $\Delta$ is depicted in Figure \ref{fig:RemCVnotV} below for the particular case that $0 < Q^2 < Q^1$. Note that for every $(p,q) \in \Delta$ it holds that $0 < p < \tfrac{Q^2}{Q^2+1}<1$ and $0 < q < \tfrac{Q^1Q^2}{Q^1+Q^2+1} <Q^2 < Q^1$.
	
\begin{figure}[h]
\centering
\begin{tikzpicture}[scale=1.5]
\def\xplusone{5}
\def\gammaval{3}
\def\intrsect{1.6667}
	
\draw[-latex] (0,0) -- (7,0) node[right] {$q$};
\draw[-latex] (0,0) -- (0,4) node[above] {$p$};
	
\node[below left] at (0,0) {$0$};
\node[left] at (0,3) {$1$};
		
\draw (0,3) -- (6,3);
		
\node[below] at (\gammaval,0) {$Q^2$};
\node[below] at (\xplusone,0) {$Q^1$};
			
\node[left, blue] at (0,2.25) {$\frac{Q^2}{Q^2+1}$};
\fill[blue] (0,2.25) circle (1pt);

\fill[color=lightgray] (0,0) -- (0,2.25) -- (1.6667,1.0) -- (0,0);
\node[black] at (0.7,0.95) {$\Delta$};
		
\draw[red, thick] (0,0) -- (\xplusone,3) node[pos=1.02, above right] {$p=\frac{q}{Q^1}$};
\draw[red, dashed] (\xplusone,3) -- (\xplusone,0);
\fill[red] (\xplusone,3) circle (1pt);
		
\draw[blue, thick] (0,2.25) -- (\gammaval,0) node[pos=0.9, above right] {$p=\frac{Q^2-q}{Q^2+1}$};
\fill[blue] (\gammaval,0) circle (1pt);

\fill[black] (\intrsect,1.0) circle (1pt);
\draw[black,dashed] (\intrsect,1.0) -- (\intrsect,0);
\node[below] at (\intrsect,0) {$\frac{Q^1Q^2}{Q^1+Q^2+1}$};
		
\end{tikzpicture}
\caption{Viable price systems in Remark \ref{remark: CV implies V}.} \label{fig:RemCVnotV}
\end{figure}
		
\noindent
Although $\Delta \neq \varnothing$ for all output levels $Q^1, Q^2 >0$, it is clear that for decreasing $k_2$-output levels the viable price set $\Delta$ shrinks. In particular, for every output level $Q^2>0$ and every price level $p \in \left[ \tfrac{Q^2}{Q^2+1} , 1 \right]$ there does not exist any input price $q>0$ for which $(p,q)$ is viable. Hence, the complete viability property is not valid for this production system as depicted in Figure \ref{fig:RemCVnotV}. Moreover, for every $p \in \left( \tfrac{Q^2}{Q^2+1} , 1 \right]$ it holds that $I_{k_2} (p,q) = Q^2 - q - (Q^2+1) p <0$ for every $q \geqslant 0$, so that the weak complete viability property fails as well. This proves that a (weakly) viable production system need not be (weakly) completely viable.
\end{remark}

\noindent
Our first characterisation of complete viability is stated next. 

\begin{proposition} \label{prop:CV-WCV}
	Let $\mathbb S = \langle N, \zeta , \gamma \rangle$ be a structured production system. Then, $\mathbb S$ is completely viable if and only if $\mathbb S$ is weakly completely viable and coherent.
\end{proposition}

\noindent
A proof of Proposition \ref{prop:CV-WCV} is provided in Appendix \ref{sec: appB1} of this paper. 

Example \ref{ex:Coherence} shows that we cannot dispense with the coherence assumption in Proposition \ref{prop:CV-WCV} because the described production system $\mathbb{S}$ is not coherent, and hence it is neither viable nor completely viable (Theorem \ref{thm:ViabilityStructural} and Remark \ref{remark: CV implies V}). On the other hand, it is weakly completely viable and a fortiori weakly viable (Remark \ref{rem: viable vs coherent}).

\paragraph{Input restrictions in production processes}

We can link the viability of the production system to feedback loops in the professional production system $\zeta$. In particular, final consumption goods cannot act as intermediate inputs in the production of certain goods. The first property imposes this to be the case for \emph{all} goods, while the second property imposes that for consumption goods only. The next definition formalises these ideas.

\begin{definition}
	Let $\mathbb S = \langle N, \zeta , \gamma \rangle$ be a structured production system.
	\begin{numm}
		\item The production system $\mathbb S$ satisfies the \textbf{restricted input condition} (RIC) if every good is produced without the input of any consumption good, i.e., for every good $k \in L$ and every consumption good $h \in L_c$ it holds that $y^k_h =0$. 
		\item The production system $\mathbb S$ satisfies the \textbf{weak restricted input condition} (WRIC) if every consumption good is produced without the input of any consumption good, i.e., for all consumption goods $k,h \in L_c$ it holds that $y^k_h =0$.
	\end{numm}
\end{definition}

\noindent
We note that, trivially, the restricted input condition RIC implies the weak restricted input condition WRIC, whereas the converse might not be true as the next example shows. 

\begin{example} \label{ex:notRIC}
	Consider the production system $\mathbb S$ with two consumables, denoted as $k_1$ and $k_2$, respectively and one intermediate good $k_3$. We again use a matrix representation for the professional production system $\zeta \colon$
	\[
	Z =
	\begin{pmatrix}
	Q^1 & 0 & -1 \\
	0 & Q^2 & 0 \\
	-1 & 0 & 1
	\end{pmatrix}
	\qquad \mbox{where } Q^1=1+x, \,\,\mbox{with}\,\,Q^2,x >0.
	\]
	Again each profession is assumed by a single agent, leading to a net productive output $\omega = (x,Q^2)^\top \gg 0$.  Note that the production system $\mathbb{S}$ satisfies WRIC but not the restricted input condition RIC because the consumption good $k_1$ is used as an input for the production of the intermediate good $k_3$. 

\begin{figure}[h]
	\centering
	\begin{tikzpicture}[scale=1.25]
	\def\xplusone{6}
	\def\point{2}
	
	\fill[lightgray] (0,0) -- (\point,3) -- (\xplusone,3) -- (0,0) -- cycle;
	
	\draw[-latex] (0,0) -- (8,0) node[right] {$q$};
	\draw[-latex] (0,0) -- (0,4) node[above] {$p$};
	
	\node[below left] at (0,0) {$0$};
	\node[left] at (0,3) {$1$};
	
	\draw[thick] (0,3) -- (7,3);
	
	\draw[red, thick] (0,0) -- (\point,3) coordinate (redtop);
	\draw[red, dashed] (redtop) -- (\point,0) node[below] {$1$};
	\fill[red] (redtop) circle (2pt);
	
	\draw[blue, thick] (0,0) -- (\xplusone,3) coordinate (bluetop);
	\draw[blue, dashed] (bluetop) -- (\xplusone,0) node[below] {$x+1$};
	\fill[blue] (bluetop) circle (2pt);
	
	\node[black, scale=1.5] at (2.75,2) {$\Delta$};
	\node[above] at (bluetop) {$p = \tfrac{q}{x+1}$};
	\node[above] at (redtop) {$p=q$};	
	\end{tikzpicture}
	\caption{Viable price system analysis for Example \ref{ex:notRIC}.} \label{fig:notRIC}
\end{figure}

\noindent
From the graphical representation in Figure \ref{fig:notRIC}, it is clear that the production system $\mathbb S$ is completely viable. Hence, this example shows that there exist completely viable structured production systems that do not satisfy the restricted input condition RIC. 
\end{example}

\subsection{Structural analysis of complete viability}\label{sec: Structural analysis of complete viability}

We now observe that RIC imposes a structural property on the matrix representation of the professional production system. 

\begin{remark}\label{rem: Z block matrix}
	Under the restricted input condition RIC, the matrix $Z$ representing the professional production system $\zeta$ can be written as a block-matrix 
	\[
	Z=
	\begin{pmatrix}
	Q_c&B \\
	0&Z_p			
	\end{pmatrix}
	\] 
	where 
	\begin{itemize}
		\item[] $Q_c$ is the positive $(\ell_c \times \ell_c)$-diagonal matrix $Q_c=(\delta_{kh}Q^k)$  where $\delta_{kh}$ is the Kronecker delta; 
		\item[] $0$ is the $(\ell_p\times \ell_c)$- null matrix; 
		\item[] $B=(b_{kh})$ is a non-positive $(\ell_c \times \ell_p)$ matrix with $b_{kh}=-y^k_h\leqslant 0$, and
		\item[] $Z_p$ is the $(\ell_p\times \ell_p)$- $\mathcal{Z}^+$-matrix with $z_{kk}=Q^k>0$ and $z_{kh}=-y^k_h\leqslant 0$ for $k \neq h$.
	\end{itemize}
	Note that some columns of $B$ may be null. However, if $\mathbb S$ is in addition coherent, not all columns of $B$ can be null. Indeed, if $B=0$, the structured production hypothesis (\ref{eq:StructuredEconomy}) implies that $Z_p^\top n_p =0$, where $n_p \gg 0$ denotes the vector of professional class sizes of the intermediate good professions, so that $\det Z = \det Q_c \cdot \det Z_p = 0$, contradicting coherence by Theorem \ref{thm:ViabilityStructural}.
\end{remark}

\noindent
The next theorem establishes that the weakly complete viability property of a structured production system is related to these restricted input conditions. We show that the weakly complete viability condition stands exactly between the two restricted input conditions introduced here. 

\begin{theorem} \label{thm:CompleteViable}
	Let $\mathbb S = \langle N, \zeta , \gamma \rangle$ be a structured production system. Then:
	\begin{abet}
		\item If the structured production system $\mathbb S$ satisfies the restricted input condition RIC, then $\, \mathbb S$ is weakly completely viable.
		\item If the structured production system $\mathbb S$ is weakly completely viable, then $\, \mathbb S$ satisfies the weak restricted input condition WRIC.
	\end{abet}
\end{theorem}

\noindent
A proof of Theorem \ref{thm:CompleteViable} is presented in Appendix \ref{sec: app B2} of this paper.

\medskip\noindent
Two counterexamples demonstrate that the converses of both parts (a) and (b) of Theorem \ref{thm:CompleteViable} fail. Example \ref{ex:notRIC} provides a weakly completely viable system that does not satisfy RIC, while Example \ref{ex:notWRIC} below shows that WRIC does not imply weak complete viability.

\begin{example}\label{ex:notWRIC}
	Consider the production system $\mathbb{S}$ with two consumables, denoted by $k_1$ and $k_2$, and two intermediate goods, denoted by $k_3$ and $k_4$, respectively. We again use a matrix representation for the professional production system $\zeta$:
	\[Z =
	\begin{pmatrix}
	4 & 0 & -1 &-1\\
	0 & 2 & 0 &-1\\
	-1 & 0 & 1 &0\\
	-1&-1&0&2
	\end{pmatrix}\]
	Again, each profession is assumed by a single agent, leading to a net productive output $\omega = (2,1)^\top \gg 0$. Note that the production system $\mathbb{S}$ satisfies WRIC but not RIC.\footnote{This is because the consumption good $k_1$ is used as an input for the production of both intermediate goods, whereas the consumption good $k_2$ is used as an input for the production of the intermediate good $k_3$ only.} 
	\\
	Furthermore, since $\det Z = 7 \neq 0$, Theorem \ref{thm:ViabilityStructural} ensures that $\mathbb{S}$ is coherent and viable. We now show that $\mathbb{S}$ is not weakly completely viable. To this end, let $p \in [0,1]$ represent the competitive market price of $k_1$, which implies that the competitive market price of consumable $k_2$ is $1-p$. Furthermore, let $q_1, q_2 \geqslant 0$ stand for the prices of the intermediate inputs $k_3$ and $k_4$, respectively. The income equations are derived as follows:
	\begin{align*}
	I_{k_1} (p,q_1,q_2) & = 4p-q_1-q_2 \\
	I_{k_2} (p,q_1,q_2) & = 2(1-p)-q_2\\
	I_{k_3} (p,q_1,q_2) & = -p+q_1\\
	I_{k_4} (p,q_1,q_2) & = -p-(1-p)+2q_2=-1+2q_2.
	\end{align*}
	Let $p^* = \frac{7}{8}$. Note that imposing weak viability (i.e., $I_{k_i} (p^*, q_1,q_2) \geqslant 0$ for all $i$) to find a pair $(q^*_1, q^*_2)$ such that $(p^*, q^*_1, q^*_2) \in \Delta'$ leads to the following contradiction: $\frac{1}{2} \leqslant q^*_2 \leqslant \frac{1}{4}$. Hence, $\mathbb{S}$ satisfies WRIC but is not weakly completely viable,  implying that the converse of Theorem \ref{thm:CompleteViable}(b) does not hold.
\end{example}

\noindent 
Due to Proposition \ref{prop:CV-WCV} and Theorem \ref{thm:CompleteViable}, we conclude that in coherent structured production systems, complete viability stands exactly between the two restricted input conditions. 

\begin{corollary}\label{cor: completeviable and RIC}
	Let $\mathbb S = \langle N, \zeta , \gamma \rangle$ be a coherent structured production system. Then:
	\begin{abet}
		\item If the structured production system $\mathbb S$ satisfies the restricted input condition, then $\, \mathbb S$ is completely viable.
		\item If the structured production system $\mathbb S$ is  completely viable, then $\, \mathbb S$ satisfies the weak restricted input condition.
	\end{abet}
\end{corollary}

\noindent
We conclude with some remarks. 

First, we note that the structured production system considered in Example \ref{ex:Coherence} is not coherent and hence not viable, but satisfies the restricted input condition RIC. Moreover, we have already noted that $\Delta = \varnothing$, whereas $\Delta' \neq \varnothing$ (see Remark \ref{rem: viable vs coherent}). Therefore, the example shows that the RIC does not imply viability, even though weak viability is satisfied, confirming the essential role of the coherence assumption in Corollary \ref{cor: completeviable and RIC}.  

 Second, the structured production system described in Remark \ref{remark: CV implies V}  is coherent and viable, but it violates the weak restricted input condition because consumable $k_1$ is used as an input in the production of consumable $k_2$. By Theorem \ref{thm:CompleteViable}(b), this production system therefore fails to satisfy the weak complete viability property as well---as was also shown directly in Remark \ref{remark: CV implies V}. Therefore, the example demonstrates that viability implies neither WRIC nor, a fortiori, weak complete viability.  

Therefore, even though viability does not imply weak complete viability, and vice versa, taken together they characterise complete viability, as shown in Proposition \ref{prop:CV-WCV}.

\section{A framework for further development}

This paper's central contribution lies in developing the concept of viability---the requirement that all producers earn positive incomes---as a prerequisite for meaningful economic participation in a structured production system and as the foundation for subsequent equilibrium analysis.

Our analysis yields several key theoretical insights. We demonstrate that viability of a structured production system is intimately connected to its structural properties. Theorems \ref{thm:ViabilityStructural} and \ref{thm:CompleteViable} provide tractable structural conditions on the production system for determining when price systems can sustain all productive activities.

\subsection{Toward a three-stage general equilibrium conception}

The viability analysis presented here forms the foundation for a comprehensive three-stage equilibrium conception. Unlike existing production network models that either focus on competitive equilibria with homogeneous price mechanisms \citep{BaqaeeFarhi2020} or analyse bilateral bargaining between buyers and suppliers in isolation \citep{Alviarez2026}, our framework explicitly accommodates the coexistence of competitive and bilateral pricing within a unified equilibrium structure. 

Our second paper \citep{GillesPesce-SP-2} introduces a general equilibrium concept that combines viable price systems with market clearing in consumption good markets. This development faces unique challenges because the pricing of intermediate goods lacks guidance through competitive markets. 

This particular general equilibrium concept ensures not only that consumption good markets clear at viable prices, but also that intermediate good allocations are consistent with the production requirements encoded in the production technology represented by the system. This involves extending classical general equilibrium theory to accommodate the nature of production chains and the absence of markets for intermediate inputs. In particular, \citet{GillesPesce-SP-2} establish the existence of general equilibrium and the two welfare theorems for economies with a completely viable structured production system.

A third paper \citep{GillesPesce-SP-3} endogenises intermediate good price formation through explicit bargaining mechanisms. This is developed within the context of a structured production system that is viable as well as in general equilibrium. We note that endogenous intermediate good pricing requires the further specification of structured production systems as \emph{production networks}. 

\subsection{Some final remarks: Broader implications}

Our research program offers a fresh perspective on production economies that bridges Leontief-Sraffian production theory with modern network economics. By explicitly modelling the distinct roles of consumption and intermediate goods, we capture essential features of real production systems obscured in traditional approaches. The framework provides tools for analysing how production network structure affects income distribution, price formation, and economic stability. As such the framework lends itself to the study of causes of inequality. 

Future extensions might incorporate dynamic elements, examining how production networks evolve when non-viable producers exit and new production technologies emerge, which is not captured in the static production systems considered here. The extended framework could also accommodate multiple production technologies per good, incomplete information about production processes, or strategic behaviour beyond bargaining.

The viability concept introduced here has immediate policy relevance. Understanding which production system structures guarantee viable price systems informs industrial organisation, supply chain design, and economic development strategies. For instance, our results suggest that production networks organised to minimise consumption goods serving as intermediate inputs exhibit greater structural stability. The conditions linking input restrictions to viability suggest that certain production architectures are inherently more robust to price fluctuations.


\newpage
\singlespace
\bibliographystyle{econometrica}
\bibliography{RPDB}

\newpage
\appendix

\section*{Appendices}

\singlespace

\section{Proofs of Section \ref{sec:Viability}}\label{sec: app A}

\subsection{Proof of Theorem \ref{thm:ViabilityStructural}}\label{sec: app A1}

We show $\text{(iii)}\Rightarrow\text{(ii)}\Rightarrow\text{(i)}\Rightarrow\text{(iii)}$.

\paragraph{(iii) implies (ii)} \ \
\\
Assume that there exists some viable price system $(p,q) \in \Delta$ for $\mathbb S$. \\
Suppose to the contrary that $\mathbb S$ is \emph{not} coherent. Then $\mathbb S$ admits some conversion cycle. By Definition \ref{def:Coherence} this implies there exists some $\alpha = (\alpha_1, \ldots , \alpha_\ell ) \in \mathbb R^\ell_+ \setminus \{ 0 \}$ with $\alpha \leqslant n$ such that  $Z^\top \alpha =0$. \\
Let $C = \{ k \in L \mid \alpha_k >0 \}$ be the set of goods that are represented in this conversion cycle. \\
Now, since $\sum_{k \in C} \alpha_k \, \zeta (k) =0$ we conclude that for $(p,q) \colon$
\begin{equation} \label{eq:CoherenceProof}
\sum_{k \in C} \alpha_k \, I_k (p,q) = \sum_{k \in C} \alpha_k \, (p,q) \cdot \zeta (k) = (p,q) \cdot \sum_{k \in C} \alpha_k \, \zeta (k) =0
\end{equation}
On the other hand, since $(p,q) \in \Delta$ it holds that $I_k (p,q) = (p,q) \cdot \zeta (k) >0$ for every $k \in C$, and hence $\sum_{k \in C} \alpha_k \, I_k (p,q) >0$. But this contradicts (\ref{eq:CoherenceProof}), showing the assertion.

\paragraph{(ii) implies (i)} \ \ 
\\
Let $\mathbb S$ be coherent and assume to the contrary that $\det Z = 0$. Define
\[
\beta \, :=\, \max_{k \in L} Q^{k} \, >\, 0,
\qquad \mbox{and} \qquad
B = (b_{kh}) \, := \,
\begin{cases}
\beta - Q^{k} \geqslant 0 & \text{if } k = h,\\
y^k_{h} \geqslant 0    & \text{if } k \neq h.
\end{cases}
\]
Then it follows that $Z= \beta I - B$, with $B \geqslant 0$. Hence,
\[
Z^\top = \left( \beta I - B \right)^\top = \beta I - B^\top \qquad \mbox{with } B \geqslant 0 \mbox{ and } \beta >0 .
\]
Furthermore, by the structured production hypothesis  \eqref{eq:StructuredEconomy}, $Z^\top n>0$ (strictly positive in the consumption good components and zero in the intermediate good components) where $n\gg0$. From Lemma \ref{Fact 1}(2) we deduce that $\rho (B^\top ) \leqslant \beta$. \\
Now, $\det Z = 0$ implies that $\det(\beta I-B) = 0$. Hence, $\beta >0$ is an eigenvalue of~$B$, so by definition of spectral radius $\rho(B)\geqslant \beta$, and therefore $\rho (B^\top ) = \rho (B) \geqslant \beta$. \\
Hence, we conclude that $\rho (B^\top) = \beta$. \\[1ex]
From Lemma \ref{Fact 1}(1b), there exists some $x\geqslant 0$ with $x \neq 0$, such that
\[
B^{\top}x \, =\,  \rho(B^{\top})\,x \, =\,  \beta\,x,
\]
and, therefore, that 
\begin{equation}\label{eq:y-eps-xi}
Z^\top x = (\beta I-B^{\top})x = 0. 
\end{equation}
Since $x >0$, the index set $ \Gamma = \{ k \in L \mid x_k >0 \} \neq \varnothing$. Now, we can define $\alpha \in \mathbb R^{\ell}_+\setminus \{0\}$ by
\[
\alpha \, =\,  \varepsilon\,x >0,\,\,\mbox{with } \varepsilon = \min_{k \in   \Gamma} \frac{n_k}{x_k} >0 .
\]
Then $\alpha > 0$ as well as $\alpha \leqslant n$, because
\[
\alpha_{k} \, =\, 
\begin{cases}
\varepsilon\cdot 0 = 0 < n_{k} & \text{if } k\notin \Gamma ,\\
\varepsilon\cdot x_{k} \leqslant \dfrac{n_{k}}{x_{k}}\cdot x_{k} = n_{k}
& \text{if } k \in\Gamma.
\end{cases}
\]
Moreover, by \eqref{eq:y-eps-xi},
\[
Z^{\top} \alpha \, {=}\,  Z^{\top}(\varepsilon x) \, =\,  \varepsilon\, Z^{\top}x \, =\,  \varepsilon\cdot 0 \, =\,  0.
\]
This contradicts the assumption that $\mathbb S$ is coherent, showing that (ii) indeed implies (i).

\paragraph{(i) implies (iii)} \ \ \\
We know that $\det Z\neq 0$ and want to show that $\mathbb S$ is viable, i.e., by Remark \ref{remark:Viable}$(ii)$ we intend to show that
\begin{equation} \label{eq:star}
\mbox{There exists some } x\gg 0 \quad\text{such that}\quad Zx \gg 0.
\end{equation}
This follows from the characterisation of non-singular $\mathcal{M}$-matrices (see \citet{Berman1994}). We provide the proof for the sake of completeness.
\\
Define the diagonal net output quantity matrix by
\[
Q \, :=\,  \mbox{diag} \, \left( Q^1, \ldots ,Q^\ell \right) = (\delta_{kh}\,Q^{k}) \, =\, 
\begin{pmatrix}
Q^{1} & 0 & \cdots & 0 \\
0 & Q^{2} & \cdots & 0 \\
\vdots & \vdots & \ddots & \vdots \\
0 & 0 & \cdots & Q^{\ell}
\end{pmatrix}. 
\]
Since $Q^{k}>0$ for all $k=1,\ldots,\ell$, $Q$ is not singular and  $\det Q = \prod_{k=1}^{\ell} Q^{k} > 0$. Hence, $Q^{-1}$ exists and is given by
\[
Q^{-1} \, =\,  \Bigl(  \tfrac{\delta_{kh}}{Q^{k}} \Bigr) \, =\, 
\begin{pmatrix}
\tfrac{1}{Q^{1}} & 0 & \cdots & 0 \\
0 & \tfrac{1}{Q^{2}} & \cdots & 0 \\
\vdots & \vdots & \ddots & \vdots \\
0 & 0 & \cdots & \tfrac{1}{Q^{\ell}}
\end{pmatrix}
\,\,\mbox{and}\,\,Q^{-1} > 0.
\]
Next, let
\[
B \, :=\,  Q - Z \, =\, 
\begin{pmatrix}
0 & y^1_{2} & \cdots & y^1_{\ell} \\
y^2_{1} & 0 & \cdots & y^2_{\ell} \\
\vdots & \vdots & \ddots  & \vdots  \\
y^{\ell}_1 & y^{\ell}_2 & \cdots & 0
\end{pmatrix}
\, =\,  (b_{kh}),\quad \mbox{where }
b_{kh} \, =\, 
\begin{cases} 
0\geqslant 0, & \text{if } k=h\\ 
y^k_{h}\geqslant 0, & \text{if } k\neq h. 
\end{cases}
\]
Therefore, it follows that
\begin{equation}\label{eq:Z-Q-B}
Z \, =\,  Q - B, \quad \mbox{with }\, Q> 0\,\mbox{and}\,  B\geqslant 0.
\end{equation}
We can now define $A \, :=\,  Q^{-1} B$. Note that $A \geqslant 0$ and that
\begin{equation}\label{eq:Q(I-A)=Z}
Q(I-A) \, =\,  Q - QA \, = \,  Q - Q(Q^{-1}B) \, =\,  Q - (QQ^{-1})B \, =\,  Q - B \, \stackrel{\eqref{eq:Z-Q-B}}{=}\,  Z.
\end{equation}
Since $Q$ is diagonal and $Q^k>0$ for all $k=1, \ldots, \ell$, if the modified problem
\begin{equation}\label{eq:starstar}
\mbox{There exists some } x\gg 0 \quad\text{such that}\quad (I-A)x \gg 0
\end{equation}
holds, then \eqref{eq:star} also holds.  Thus our modified goal is to show \eqref{eq:starstar}.
\\[1ex]
From \eqref{eq:StructuredEconomy} and \eqref{eq:Z-Q-B} we conclude that
\begin{equation}\label{eq:BT-leq-Q}
0 \, \leqslant \,  Z^{\top} n \, =\,  (Q-B)^{\top} n \, =\,  Q^{\top} n - B^{\top} n \, =\,  Qn - B^{\top} n, \qquad \mbox{implying } B^{\top} n \, \leqslant \,  Q n.
\end{equation}
Denote $t = Q \, n = \left( n_1 Q^1 , \ldots , n_\ell Q^\ell \right)^\top \gg 0$. Then by the definition of $A$ we have that
\[
A^{\top} t \, =\,  (Q^{-1}B)^{\top} t \, =\,  \bigl(B^{\top}(Q^{-1})^{\top}\bigr) t  \, =\,  (B^{\top} Q^{-1})\, t  =\,  B^{\top}(Q^{-1}Q)\,n \, =\,  B^{\top} n\, \stackrel{\eqref{eq:BT-leq-Q}}{\leqslant}\,  Qn  \, = \,  t .
\]
Also $A^{\top} t \geqslant 0$ since $t \gg 0$ and $A \geqslant 0$.  Therefore,
\begin{equation}\label{eq:sandwich}
0\cdot t \, \leqslant \,  A^{\top} t \, \leqslant \,  1 \cdot t .
\end{equation}
By Lemma \ref{Fact 1}(1e-f) applied to \eqref{eq:sandwich}, we deduce that $0 \leqslant \rho(A^{\top}) \leqslant 1$. Since $\rho(A) = \rho(A^{\top})$, it follows that
\[ 0 \, \leqslant \,  \rho(A) \, \leqslant \,  1.\]
From (\ref{eq:Q(I-A)=Z}), we have that $\det Z = \det Q \cdot \det(I-A)$. Since $\det Q > 0$ and $\det Z \neq 0$, we get $\det(I-A)\neq 0$. This actually implies that $\rho(A) < 1$.\footnote{Indeed, if $\rho(A) = 1$, then $1$ is an eigenvalue of $A$ (see Lemma \ref{Fact 1}(1a)),  giving $\det(I-A)=0$, which is a contradiction. Therefore, $\rho(A) \, <\,  1.$}

This implies further by Lemma \ref{Fact 1}(1c) that $A$ is a convergent matrix. Therefore, $(I-A)^{-1}$ exists and it is given by
\[
(I-A)^{-1} \, =\,  \sum_{m=0}^{\infty} A^{m} \, \geqslant \,  0.
\]
Let $v\gg 0$ and define
\begin{equation}\label{eq:x-def}
x \, :=\,  (I-A)^{-1}\,v \geqslant 0 .
\end{equation}
Write $(I-A)^{-1} = (a_{kh})\geqslant 0$ then, by \eqref{eq:x-def}, $x_{k} \, =\,  \sum_{h=1}^{\ell} a_{kh}\,v_{h} \, \geqslant \,  0$ for every $k \in L$.  \\
If $x_{k}=0$ for some $k$, then since $a_{kh}\geqslant 0$ and $v_{h}>0$, we conclude that
\[
\sum_{h=1}^{\ell} a_{kh}\,v_{h} = 0 \qquad \mbox{implying that }  a_{kh} = 0 \mbox{ for all } h.
\]
This would mean $(I-A)^{-1}$ has a null row, which is impossible since $\det(I-A)^{-1}\neq 0$.  Therefore, $x \gg 0$. Furthermore,
\[
(I-A)x \, \stackrel{\eqref{eq:x-def}}{=}\,  (I-A)(I-A)^{-1}\,v \, =\,  I\,v \, =\,  v \, \gg\,  0.
\]
Hence \eqref{eq:starstar} holds, and this completes the proof of the assertion.\\

\vskip 0.2cm

\subsection{Proof of Proposition \ref{prop:DeltaProps}}\label{sec: app A2}

		\emph{Proof of (a):} Let $P= \bar S \times \mathbb R^{\ell_p}_+$. For each \(k\in L\), define the function $	I_k:P\to \mathbb{R}$, where
		\[
		I_k(p,q)=
		\begin{cases}
		p_kQ^k-(p,q)\cdot y^k, & k\in L_c,\\[1ex]
		q_kQ^k-(p,q)\cdot y^k, & k\in L_p.
		\end{cases}
		\]
		Each \(I_k\) is affine and hence continuous. We have
	
		\[
		\Delta'=\bigcap_{k\in L}\{(p,q)\in P: I_k(p,q)\geqslant 0\}.
		\]
		
\noindent	For every \(k\in L\), $\{(p,q)\in P:I_k(p,q)\geqslant 0\}	=I_k^{-1}([0,\infty))$
		is closed because \(I_k\) is continuous and $P$ is closed. Hence \(\Delta'\) is  the intersection of closed 	half-spaces and $P$.  
	We next show that $\Delta'$ is bounded; the Minkowski-Weyl Theorem (see Section 2.1) then implies that $\Delta'$ is a (compact) polytope. \\
	First, we claim that coherence implies that every intermediate good in $L_p$ is a direct or indirect input to the production of some consumption good. Suppose to the contrary that the set $T \subseteq L_p$ of intermediate goods that are neither direct nor indirect inputs to the production of any consumption good is non-empty. Then every user of a good $j \in T$ is itself a producer of a good in $T \colon$ if $y^k_j >0$ for some $k \in L \setminus T$, then $j$ would be a direct or indirect input to the production of some consumption good through $k$. \\
	Hence, $z_{kj} =0$ for all $k \in L \setminus T$ and $j \in T$, so that, after a simultaneous reordering of rows and columns, $Z$ is block triangular and $\det Z = \det Z_{TT} \cdot \det Z_{T^cT^c}$, where $Z_{TT}$ denotes the submatrix of $Z$ with rows and columns in $T$ and $Z_{T^cT^c}$ denotes the submatrix for $T^c = L \setminus T$. Moreover, the structured production hypothesis (\ref{eq:StructuredEconomy}) requires the total net output of every intermediate good $j \in T$ to vanish and, since only producers of goods in $T$ use these goods, $\left( Z_{TT} \right)^\top n_T =0$, where $n_T \gg 0$ collects the professional class sizes of the professions in $T$. Therefore, $\det Z_{TT} =0$ and, consequently, $\det Z =0$, contradicting the coherence of $\mathbb S$ through Theorem \ref{thm:ViabilityStructural}. This shows the claim. \\
	Now, take any $(p,q) \in \Delta'$. Since $p \in \bar S$, the consumption good price vector $p$ is bounded by $(1, \ldots ,1)$. For $m \in \{ 0,1,2, \ldots \}$ define $D_m \subseteq L_p$ recursively by letting $D_0$ be the set of intermediate goods that are direct inputs to the production of some consumption good, and $D_{m+1} = D_m \cup \{ j \in L_p \mid y^h_j >0 \mbox{ for some } h \in D_m \}$. By the claim above, there is some finite $\bar m$ such that $D_{\bar m} = L_p$. If $j \in D_0$, then $y^k_j >0$ for some $k \in L_c$, and $I_k (p,q) \geqslant 0$ implies that $q_j \leqslant \tfrac{p_k Q^k}{y^k_j} \leqslant \tfrac{Q^k}{y^k_j}$. Proceeding inductively, if $j \in D_{m+1} \setminus D_m$, then $y^h_j >0$ for some $h \in D_m$, and $I_h (p,q) \geqslant 0$ implies that $q_j \leqslant \tfrac{q_h Q^h}{y^h_j}$, where $q_h$ is bounded by the induction hypothesis. Since all of these bounds depend only on the output levels $Q^k$ and the input requirements $y^k_h$, we conclude that $\Delta'$ is indeed bounded. 
	
	\medskip\noindent
	\emph{Proof of (b):} With reference to the construction introduced in the proof of (a), we note that 
	\[
	\Delta=\bigcap_{k\in L}\{(p,q)\in P: I_k(p,q)>0\},
	\]
whereas \(\Delta'\) is closed and contains \(\Delta\). Hence, $\operatorname{cl}(\Delta)\subseteq \Delta'.$\\
	Take any \((p,q)\in \Delta'\). Then $	I_k(p,q)\geqslant 0$ for all $ k\in L$.	
	Since $\mathbb S$ is coherent and hence viable by Theorem \ref{thm:ViabilityStructural}, $\Delta \neq \varnothing$; choose $	(\bar p,\bar q)\in \Delta$.	Hence
	 $	I_k(\bar p,\bar q)>0$ for all $ k\in L$.	
	For \(t\in(0,1]\), define $	(p_t,q_t)
	=(1-t)(p,q)+t(\bar p,\bar q)$. Because \(P= \bar S \times \mathbb R^{\ell_p}_+\) is convex, $(p_t,q_t)\in P$.	Since each \(I_k\) is affine,
$
	I_k(p_t,q_t)
	=(1-t)I_k(p,q)+tI_k(\bar p,\bar q).
$
	Now \(I_k(p,q)\geqslant 0\) and \(I_k(\bar p,\bar q)>0\), so for every
	\(t>0\),
	\[
	I_k(p_t,q_t)>0.
	\]
	Therefore $
	(p_t,q_t)\in \Delta$ for all $t\in(0,1],
$ and 
	$	(p_t,q_t)\to(p,q)$ as $ t\downarrow0$.
	Thus \((p,q)\in\operatorname{cl}(\Delta)\), which means that
	$
	\Delta'\subseteq \operatorname{cl}(\Delta),
$ and hence
	\[
	\Delta'=\operatorname{cl}(\Delta).
	\]
	Since $\Delta' = \mathrm{cl} \, \Delta$, it now follows from Theorem 6.3 of \citet{Rockafellar1970} that $\mathrm{ri} \, \Delta' = \mathrm{ri} \, \left( \mathrm{cl} \, \Delta \right) = \mathrm{ri} \, \Delta$.

\subsection{Algorithmic Proof of Proposition \ref{prop: acyclic implies viable}}\label{sec: app Acyc}

Let $Z$ be the $( \ell \times \ell)$ matrix representation of the professional production system $\zeta$, which  is of class $\mathcal Z^+$.  \\
To simplify the notation, we reformulate Definition \ref{def:Acyclic} of cycle and acyclicity  as follows. 

A \emph{cycle} in $Z$ is a sequence $x =(x_1, \ldots , x_m)$ such that
\begin{itemize}
	\item $x_1 = z_{ij}$ for some $i,j \in L$ with $i \neq j$;
	\item For every $s = 1, \ldots ,m-1$ we have $x_s = z_{\alpha \beta}$ and $x_{s+1} = z_{\beta \gamma}$ for some $\alpha , \beta ,\gamma \in L$ with $\alpha \neq \beta \neq \gamma$;
	\item and $x_m = z_{ti}$ for some $t \in L \setminus \{ i \}$.
\end{itemize}
By Remark \ref{rem:Acyclicity}, acyclicity of $\zeta$ means that every cycle $x$ in $Z$ is a \emph{zero cycle}, i.e., $\prod^m_{s=1} x_s =0$. Any $(\ell \times \ell )$-matrix that satisfies this property can be referred to as \emph{acyclic}. Hence, $Z$ is acyclic by hypothesis. \\[1ex]
Let $c = (c_1, \ldots ,c_\ell)^\top \gg 0$ be a positive vector in $\mathbb R^\ell$ and consider the system of equations $Zx =c$. We introduce an algorithmic process to apply iterated Gaussian elimination to the system of equations $Zx=c$. Our goal is to find a price system $(p,q)\in (\bar{S}\cap \mathbb{R}^{\ell_c}_{++})\times \mathbb{R}^{\ell_p}_{++}$ such that $Z(p,q)\gg0$ (see Remark \ref{remark:Viable}$(ii)$). \\
Denote for every $i \in L$ the $i$-th row of $Z$ by $\mathbf{z}_i = \zeta(i)^\top = (z_{i1} , \ldots , z_{i\ell})$. Now we apply the following algorithm:
\begin{description}
	\item[Step 1:] Given $Z$, since $z_{11} = Q^1 >0$, we can apply Gaussian elimination to transform the rows of $Z$ to eliminate its first column (except $z_{11}$), also modifying the vector $c$. This constructs a modified $(\ell \times \ell)$ matrix $Z^1$ with rows denoted as $\mathbf{z}^1_1 , \ldots , \mathbf{z}^1_\ell$ and a vector $c^1 \in \mathbb R^\ell$ where \\[1ex] 
	$\triangleright \qquad \mathbf{z}^1_1 = \mathbf{z}_1$; \\[1ex] 
	$\triangleright \qquad \mathbf{z}^1_i = \mathbf{z}_i - \frac{z_{i1}}{Q^1} \mathbf{z}_1$, for any $i = 2, \ldots , \ell$; \\[1ex] 
	$\triangleright \qquad c^1_1 = c_1 >0$; and \\[1ex] 
	$\triangleright \qquad c^1_i = c_i - \frac{z_{i1}}{Q^1} c_1 \geqslant c_i >0$, for any $i = 2, \ldots , \ell$. \\[1ex] 
	Note that $z^1_{1j} = z_{1j}$ for all $j \in L$ and $z^1_{ij} = z_{ij} - \frac{z_{i1} z_{1j}}{Q^1}$ for all $i,j \in L$ with $i \neq 1$. Also, $c^1 \gg 0$. In particular, since $Z$ is acyclic and of class $\mathcal Z^+$, we have that
	\begin{align*}
	z^1_{11}&=Q^1>0,\\
		z^1_{1j} & = z_{1j} = - y^1_j \leqslant 0,\,\,\mbox{for}\, j \neq 1\\
	z^1_{i1} & = -y^i_1 - \frac{-y^i_1 Q^1}{Q^1} =0,\,\,\mbox{for}\, i \neq 1\\
		z^1_{ii} & = z_{ii}  - \frac{z_{i1} z_{1i}}{Q^1} = Q^i - \frac{0}{Q^1} = Q^i >0,\,\,\mbox{for} \, i \neq 1 \\
	z^1_{ij} & = - y^i_j - \frac{z_{i1} z_{1j}}{Q^1} = - y^i_j - \frac{y^i_1 y^1_j}{Q^1} \leqslant 0, \,\,\mbox{for}\,\, i \neq 1\,\mbox{and}\, \ j  \neq i. 
		\end{align*}
	It is obvious that, therefore, $Z^1$ is of class $\mathcal Z^+$. Moreover, \\[1ex]
	\emph{\textbf{Claim:} $Z^1$ is acyclic.} \\
	\textbf{Proof of the claim:} Consider a cycle $x =(x_1, \ldots , x_m)$ such that
	\begin{itemize}
		\item $x_1 = z^1_{ij}$ for some $i,j \in L$ with $i \neq j$;
		\item For every $s = 1, \ldots ,m-1$ we have $x_s = z^1_{\alpha \beta}$ and $x_{s+1} = z^1_{\beta \gamma}$ for some $\alpha , \beta ,\gamma \in L$ with $\alpha \neq \beta \neq \gamma$;
		\item and $x_m = z^1_{ti}$ for some $t \in L \setminus \{ i \}$
	\end{itemize}
	We next show that $\prod^m_{s=1} x_s =0$. Note that by definition of $Z^1$, for any $s\in \{1,\ldots, m\}$, there are some $\alpha , \beta \in L$ with $\alpha \neq \beta$ such that $x_s = z_{\alpha \beta} - d_{\alpha \beta}$ with $d_{\alpha \beta} = \frac{z_{\alpha 1} z_{1 \beta}}{Q^1}$ if $\alpha \neq 1$ and $d_{1 \beta} = 0$. Expanding the product $\prod^m_{s=1} x_s = \prod^m_{s=1} \left( z_{\alpha_s \beta_s} - d_{\alpha_s \beta_s} \right)$ yields a finite sum of terms, each of which is, up to a positive scalar factor, a product of off-diagonal entries of $Z$ along a closed walk: every substituted term $\frac{z_{\alpha 1} z_{1 \beta}}{Q^1}$ replaces the step from $\alpha$ to $\beta$ by the two steps from $\alpha$ to $1$ and from $1$ to $\beta$. Since $Z$ is acyclic, every such product along a closed walk vanishes. Hence, $\prod^m_{s=1} x_s =0$, showing the claim. $\qquad\lozenge$
	\item[Step k with $k \in \{ 2, \ldots , \ell-1 \}$:]
	Assume that $Z^{k-1}$ and $c^{k-1}$ have been constructed with $Z^{k-1}$ acyclic and of class $\mathcal Z^+$, and $c^{k-1} \gg 0$. We now proceed to apply Gaussian elimination on $Z^{k-1}$ with its $k$-th row, modifying $c^{k-1}$ as well. \\
	Given $Z^{k-1}$, since $z^{k-1}_{kk} = Q^k >0$, we can again apply Gaussian elimination to transform the rows of $Z^{k-1}$ to eliminate all entries below the diagonal in its $k$-th column. This constructs a modified $(\ell \times \ell)$ matrix $Z^k$ with rows denoted as $\mathbf{z}^k_1 , \ldots , \mathbf{z}^k_\ell$ where \\[1ex] 
	$\triangleright \qquad \mathbf{z}^k_i = \mathbf{z}^{k-1}_i$, for $i\leqslant k$; \\[1ex] 
	$\triangleright \qquad \mathbf{z}^k_i = \mathbf{z}^{k-1}_i - \frac{z^{k-1}_{ik}}{Q^k} \mathbf{z}^{k-1}_k$, for  $i>k$; \\[1ex] 
	$\triangleright \qquad c^k_i = c^{k-1}_i >0$, for $i\leqslant k$, and \\[1ex] 
	$\triangleright \qquad c^k_i = c^{k-1}_i - \frac{z^{k-1}_{ik}}{Q^k} c^{k-1}_k \geqslant c^{k-1}_i >0$, for any $i>k$. \\[1ex] 
	Hence, $c^k \gg 0$. Furthermore, $z^k_{ij} = z^{k-1}_{ij}$ for   $i \leqslant k$ and $j \in L$ and $z^k_{ij} = z^{k-1}_{ij} - \frac{z^{k-1}_{ik} z^{k-1}_{kj}}{Q^k}$ for   $i>k $ and $j \in L$. \\ In particular, since $Z^{k-1}$ is acyclic and of class $\mathcal Z^+$, we have that
	\begin{align*}
	z^k_{ii}&= Q^i>0, \,\,\mbox{for}\, i\leqslant k\\
	z^k_{ii} & = z^{k-1}_{ii}  - \frac{z^{k-1}_{ik} z^{k-1}_{ki}}{Q^k} = Q^i - \frac{0}{Q^k} = Q^i >0,\,\,\mbox{for}\, i >k \\
	z^k_{ij} & \leqslant 0, \,\,\mbox{for}\, i,j \in L \mbox{ with } i \neq j\,\,\mbox{and,\,in\,particular,}\\
		z^k_{ij} & =  0, \,\,\mbox{for}\, i,j \in L \mbox{ with } i>j\,\mbox{and}\, j \leqslant k 		
	\end{align*}
	This implies that $Z^k$ is of class $\mathcal Z^+$. Furthermore, \\[1ex]
	\emph{\textbf{Claim:} Since $Z^{k-1}$ is acyclic, $Z^k$ is acyclic.} \\
	\textbf{Proof of the claim:} Consider a cycle $x =(x_1, \ldots , x_m)$ in $Z^{k}$ such that
	\begin{itemize}
		\item $x_1 = z^k_{ij}$ for some $i,j \in L$ with $i \neq j$;
		\item For every $s = 1, \ldots ,m-1$ we have $x_s = z^k_{\alpha \beta}$ and $x_{s+1} = z^k_{\beta \gamma}$ for some $\alpha , \beta ,\gamma \in L$ with $\alpha \neq \beta \neq \gamma$;
		\item and $x_m = z^k_{ti}$ for some $t \in L \setminus \{ i \}$
	\end{itemize}
	We next show that $\prod^m_{s=1} x_s =0$. \\ Note that by definition of $Z^{k}$, for any $s\in \{1,\ldots, m\}$, there are some $\alpha , \beta \in L$ with $\alpha \neq \beta$ such that $x_s = z^{k-1}_{\alpha \beta} - d_{\alpha \beta}$ with $d_{\alpha \beta} = \frac{z^{k-1}_{\alpha k} z^{k-1}_{k \beta}}{Q^k}$ if $\alpha > k$ and $d_{\alpha \beta} = 0$ if $\alpha \leqslant k$. As in Step 1, expanding the product $\prod^m_{s=1} x_s$ yields a finite sum of terms, each of which is, up to a positive scalar factor, a product of off-diagonal entries of $Z^{k-1}$ along a closed walk, obtained by replacing the step from $\alpha$ to $\beta$ by the two steps from $\alpha$ to $k$ and from $k$ to $\beta$. Since $Z^{k-1}$ is acyclic, every such product along a closed walk vanishes. Hence, $\prod^m_{s=1} x_s =0$, showing the claim. $\qquad\lozenge$
\end{description}
Based on the induction process developed above, we conclude that $Z, Z^1, \ldots ,Z^{\ell-1}$ are all acyclic and of class $\mathcal Z^+$. In particular, the matrix $Z^{\ell -1}$ is acyclic as well as upper triangular:
\begin{align*}
z^{\ell -1}_{ii} &= Q^i >0,\,\,\mbox{for}\, i \in L \\[1ex]
z^{\ell -1}_{ij}& \leqslant 0,\,\,\mbox{for}\, j>i \\[1ex]
z^{\ell -1}_{ij}&=0,\,\mbox{for}\, j < i.
\end{align*}
Since $Z^{\ell -1}$ is upper triangular with a strictly positive diagonal, the system of equations $Z^{\ell -1} x = c^{\ell -1}$ admits a unique solution $x = (p,q) \in \mathbb R^\ell$, which can be computed through backward substitution. We claim that $x = (p,q) \gg 0 \colon$
\begin{itemize}
	\item The last equation of the system $Z^{\ell -1} x = c^{\ell -1}$ states that $Q^\ell x_\ell =c^{\ell -1}_\ell$, so that $x_\ell = \tfrac{c^{\ell -1}_\ell}{Q^\ell} >0$.
	\item The preceding equation $\ell -1$ states that
	\[
	Q^{\ell -1} x_{\ell -1} + z^{\ell -1}_{\ell -1, \ell} x_\ell = c^{\ell -1}_{\ell -1}, \qquad \mbox{implying } \quad x_{\ell -1} = \tfrac{1}{Q^{\ell -1}} \left[ \, c^{\ell -1}_{\ell -1} - z^{\ell -1}_{\ell -1, \ell} x_\ell \, \right] >0,
	\]
	since $z^{\ell -1}_{\ell -1, \ell}  \leqslant 0$, $c^{\ell -1}_{\ell -1} >0$ and $x_\ell >0$.
	\item Proceeding backward, we arrive at
	\[
	x_{k} = \tfrac{1}{Q^{k}} \left[ \, c^{\ell -1}_{k} - \sum_{h>k} z^{\ell -1}_{k, h} x_h \, \right] >0,\,\, \mbox{for all } k=1, \ldots , \ell .
	\]
\end{itemize}
Since Gaussian elimination preserves the solution set of a system of equations, $x$ also solves the original system, i.e., $Zx = Z(p,q) = c \gg 0$. \\
We can next normalise the identified $x= (p,q) \gg 0$ to arrive at a price system $( \tilde p , \tilde q) \in  (\bar S\cap \mathbb R^{\ell_c}_{++}) \times \mathbb R^{\ell_p}_{++}$. Furthermore, the inequality $Z (\tilde p, \tilde q) \gg 0$ is preserved in this normalisation. The row-wise statement of this inequality leads to the conclusion that 
\[
( \tilde p, \tilde q) \cdot \zeta (k) = I_k ( \tilde p, \tilde q) >0,\,\, \mbox{for every $k \in L$.}
\]
Hence, $( \tilde p, \tilde q) \in \Delta \neq \varnothing$ is a viable price system.

\section{Proofs of Section \ref{sec: Complete Viability of structured production systems price systems}}\label{sec: app B}

\subsection{Proof of Proposition \ref{prop:CV-WCV}}\label{sec: appB1}

Let $\mathbb S$ be completely viable. From Remark \ref{remark: CV implies V}, $\mathbb{S}$ is viable and hence coherent (see Theorem \ref{thm:ViabilityStructural}). We need to show that for every $p \in \bar S$ there exists some $q \in \mathbb R^{\ell_p}_+$ with $(p,q) \in \Delta'$. \\
	If $p \in \bar S \cap \mathbb R^{\ell_c}_{++}$ then by complete viability, there exists $q \in \mathbb R^{\ell_p}_{++}$ with $(p,q) \in \Delta \subseteq \Delta'$. \\
	Next, let $p \in \bar S \setminus \mathbb R^{\ell_c}_{++}$ be a price vector on the relative boundary of $\bar S$. Define a sequence $p^m \in \mathbb R^{\ell_c}_{++}$ by
	\[
	p^m_k = \frac{p_k + \tfrac{1}{m \ell_c}}{1+ \tfrac{1}{m}} \to p_k \qquad \mbox{as } m \to \infty .
	\]
	Note that for all $m$, $p^m \gg 0$ and $\sum^{\ell_c}_{k=1} p^m_k =1$. Hence, $p^m \in \bar S \cap \mathbb R^{\ell_c}_{++}$ for all $m$. \\
	Since $\mathbb S$ is completely viable, for any $m$ there exists some $q^m \in \mathbb R^{\ell_p}_{++}$ with $\left( p^m,q^m \right) \in \Delta\subseteq \Delta'$. Since $\mathbb S$ is coherent, by Proposition \ref{prop:DeltaProps}, $\Delta'$ is compact, implying that the constructed sequence $\left( p^m,q^m \right)_{m \in \mathbb N}$ admits a convergent subsequence. Hence, there exists some $q \in \mathbb R^{\ell_p}_+$ such that, along this subsequence, $\left( p^m,q^m \right) \to (p,q) \in \Delta'$. This shows that $\mathbb S$ is indeed weakly completely viable.

\medskip\noindent
To show the reverse, let $\mathbb S$ be weakly completely viable as well as coherent. Then by Theorem \ref{thm:ViabilityStructural} it is viable as well.  By viability there exists some $( p , q ) \in \Delta \subseteq \bar S \times \mathbb R^{\ell_p}_+$. In particular, by Remark \ref{remark:Viable} (i), it holds that $p \gg 0$ and $q \gg 0$. \\
	Take any $\hat p \in \bar S$ such that $\hat p \neq p$ and $\hat p \gg 0$. We now show there exists some $\hat q \in \mathbb R^{\ell_p}_{++}$ such that $(\hat p, \hat q) \in \Delta$. For that purpose identify
	\[
	\hat p_k = \min \{ \hat p_i \mid i=1, \ldots , \ell_c \} \in \left( 0, \tfrac{1}{\ell_c} \right] \qquad \mbox{and} \qquad p_h = \max \{ p_i \mid i=1, \ldots , \ell_c \} \in \left[ \tfrac{1}{\ell_c} ,1 \right) .
	\]
	Take some $\lambda$ with $0 \leqslant 1- \tfrac{\hat p_k}{p_h} < \lambda <1$; such a $\lambda$ exists, since $0 < \tfrac{\hat p_k}{p_h} \leqslant 1$. Next, define
	\begin{equation}
	\bar p = \tfrac{1}{\lambda} \, \left( \hat p - (1- \lambda) p \, \right)
	\end{equation} 
	\textbf{Claim:} $\bar p \in \bar S$ and $\bar p \gg 0$. \\[1ex]
	\emph{Proof of the claim:} First, we show that $\bar p \gg 0$. Indeed, note that for every $i \in L_c \colon$
	\begin{align*}
	\hat p_i - (1- \lambda ) p_i &= \hat p_i - p_i + \lambda p_i > \hat p_i - p_i + \left( 1 - \tfrac{\hat p_k}{p_h} \right) p_i \\[1ex]
	& = \hat p_i - \tfrac{\hat p_k}{p_h} p_i \geqslant \hat p_i - \hat p_k \geqslant 0
	\end{align*}
	since $p_i \leqslant p_h$ and $\hat p_i \geqslant \hat p_k$. \\
	Next, note also that
	\[
	\sum_{i \in L_c} \bar p_i = \tfrac{1}{\lambda} \, \left( \sum_{i \in L_c} \hat p_i - (1- \lambda) \sum_{i \in L_c} p_i \, \right) = \tfrac{1}{\lambda} \, \left( 1 - (1- \lambda) \, \right) =1
	\]
	This shows the claim. \hfill $\lozenge$ \\[1ex]
	By weak complete viability there exists some $\bar q \in \mathbb R^{\ell_p}_+$ such that $( \bar p , \bar q ) \in \Delta'$. \\
	Next, define $\hat q = \lambda \bar q + (1- \lambda ) q \gg 0$, since $\bar q \geqslant 0$, $q \gg 0$ and $0 < \lambda < 1$. \\
	Now, $( \hat p, \hat q) = \lambda ( \bar p , \bar q) + (1- \lambda ) (p , q)$ with $(\bar p , \bar q ) \in \Delta'$ and $(p , q ) \in \Delta$. In particular, $( \hat p, \hat q) \in (\bar S\cap \mathbb R^{\ell_c}_{++})  \times \mathbb R^{\ell_p}_{++}$ and by linearity of the inner product we have for every $k \in L \colon  I_k ( \hat p, \hat q) = \lambda I_k ( \bar p, \bar q ) + (1- \lambda ) I_k ( p, q ) >0$, because $\lambda \in (0,1)$, $I_k ( \bar p, \bar q ) \geqslant 0$, and $I_k ( p, q ) >0$. \\
	Hence, for every $\hat p \in \bar S$ with $\hat p \gg 0$, there exists some $\hat q \in \mathbb R^{\ell_p}_{++}$ such that $(\hat p, \hat q) \in \Delta$,  concluding the proof.

\subsection{Proof of Theorem \ref{thm:CompleteViable}}\label{sec: app B2}


\paragraph{Proof of (a):}
Suppose that $\mathbb S = \langle N, \zeta , \gamma \rangle$ satisfies the restricted input condition. Next, suppose that $\mathbb S$ is \emph{not} weakly completely viable. \\
Then there exists some $\bar p \in \bar S$ such that for every $q \in \mathbb R^{\ell_p}_+$ there is some profession $k_q \in L$ with $I_{k_q} (\bar p ,q) <0$. In particular, take $q=0$ and let $k_0$ be the corresponding profession for which $I_{k_0} (\bar p, 0) <0$. Then\footnote{Here, we decompose $y \in \mathbb R^\ell$ as $(y_{L_c} , y_{L_p})$ with $y_{L_c}$ the projection of $y$ on $\mathbb R^{\ell_c}$ and $y_{L_p}$ the projection of $y$ on $\mathbb R^{\ell_p}$.}
\[
I_{k_0} (\bar p,0) = \left\{
\begin{array}{ll}
\bar p_{k_0} \, Q^{k_0} - \bar p \cdot y^{k_0}_{L_c} \qquad & \mbox{if } k_0 \in L_c \\
- \bar p \cdot y^{k_0}_{L_c} & \mbox{if } k_0 \in L_p
\end{array}
\right.
\]
Since $I_{k_0} (\bar p, 0) <0$, in either case there exists some $h_0 \in L_c$ such that $y^{k_0}_{h_0} >0$. But this would violate the hypothesis that $\mathbb S$ satisfies the restricted input condition. This shows assertion (a).

\paragraph{Proof of (b):}
Suppose that $\mathbb S = \langle N, \zeta , \gamma \rangle$ is weakly completely viable. If $\ell_c =1$, then $\mathbb S$ trivially satisfies WRIC. Therefore, let $\ell_c \geqslant 2$.  Now, suppose to the contrary that $\mathbb S$ does \emph{not} satisfy the weak restricted input condition. We again show that a contradiction arises. \\
Indeed, since $\mathbb S$ does not satisfy the weak restricted input condition, there exists some consumption good $k \in L_c$ with profession $\zeta (k) = Q^k \, e^k - y^k$ such that consumption good input vector $y^k_{L_c} >0$. \\
Let $h \in L_c \setminus \{ k \}$ be such that $y^k_h >0$.  Since $k \in L_c$, for any price system $(p,q) \colon$
\[
I_k (p,q) = p_k \, Q^k - p \cdot y^k_{L_c} - q \cdot y^k_{L_p} \leqslant  p_k \, Q^k - p_h \, y^k_h .
\]
Take a consumption good price vector $\bar p \in \bar S$ such that $0< \bar p_i < \tfrac{y^k_h}{(\ell_c-1)(Q^k + y^k_h)}$ for all $i \neq h$ and $\bar p_h = 1- \sum_{i \in L_c \setminus \{ h \}} \bar p_i  > \tfrac{Q^k}{Q^k+y^k_h}>0$. Then we conclude that irrespective of the intermediate good price vector $q \in \mathbb R^{\ell_p}_+ \colon$
\[
I_k (\bar p,q) \leqslant  \bar p_k \, Q^k - \bar p_h \, y^k_h < \tfrac{y^k_h}{(\ell_c-1)(Q^k + y^k_h)} \, Q^k - \tfrac{Q^k}{Q^k+y^k_h} \, y^k_h = \frac{Q^k}{Q^k+y^k_h} y^k_h \left( \frac{2- \ell_c}{\ell_c -1} \right) \leqslant 0 
\]
since $\ell_c \geqslant 2$. \\
This contradicts the hypothesis that $\mathbb S$ is weakly completely viable, implying that necessarily $\mathbb S$ indeed satisfies the weak restricted input condition.

\end{document}